\documentclass[]{article}
\usepackage{graphicx}
\usepackage{braket}
\usepackage{amsmath}
\usepackage{physics}
\usepackage{bbold}
\usepackage{verbatim}
\usepackage{amssymb}
\usepackage{subfigure}
\usepackage[font=scriptsize]{caption} 
\usepackage{hyperref}
\let\Contentsline\contentsline
\renewcommand

\begin{document}
	
\title{Control with EIT: High energy charged particle detection}
\author{Aneesh Ramaswamy,$\text{ }$ Svetlana A.Malinovskaya\thanks{smalinov@gmail.com}\\
	 \textit{Department of Physics, Stevens Institute of Technology, NJ 07030, USA}}

\maketitle
\thispagestyle{empty}
\begin{abstract}
The strong non-linear optical response of atomic systems in electromagnetically induced transparency (EIT) states is considered as a means to detect the presence of small perturbations to steady states. For the 3 level system, expressions for the group velocity and group velocity dispersion (GVD) were derived and a quantum control protocol was established to account for the change in the chirp spectrum of a probe pulse when the steady state was perturbed. This was applied to the propagation of slow Cherenkov polaritons in the medium due to the passage of a train of high-energy charged particles (high energy particles). The choice of the initial steady state with focus on the slow light condition and strong narrowly confined dispersion, equated to the continuous trapping of Cherenkov polaritons in the medium along a narrow group cone, allowing for non-trivial fields to accumulate. Considering another medium prepared for the detection of the radiation, sweeping of the control field and detuning parameters in the field-atom parameter space showed the presence of optimal regions to maximize the first order perturbation in the coherences creating changes in the optical responses that modify the chirp spectra of probe pulses.
\end{abstract}

\newpage

\tableofcontents
\thispagestyle{empty}

\newpage

\section{Introduction}

\subsection{A brief review on EIT}
The use of coherent interactions between light and matter yield phenomena in which there is drastic change in the optical response function. EIT is one of those phenomenon in which a narrow transparency window with strong non-linear optical effects is achieved due to interference between quantum excitation pathways. The result is an almost negligible amplitude and for an absorptive transition that shows us Fano resonances in the optical response functions. Of prime interest is the greatly reduced absorption and drastically increased dispersion effects. Whilst its effects has analogues in CPT (Coherent population trapping), which occurs in optically thin media, and ATS (Autler-Townes Splitting); EIT is a phenomenon that involves modification of both the optical and material states of the coupled light-matter system and hence occurs in optically thick systems. Of particular interest is the study of quantum fields in EIT systems as dark state polaritons, the pseudo-particle arising from the entanglement of the propagating light with the dipole transitions.

Theoretical studies have opened up several applications including generation of non-classical atomic ensembles, high-resolution spectroscopy and reversible quantum memories in optical systems. A proposed technique by Fleischhauer and Mewes to greatly increase storage time for use in quantum memory is in adiabatically reducing the control field's strength to bring the light to a halt, effectively mapping the light's quantum state to the atomic spin ensemble and then achieve reconstruction through restoring the control field's strength \cite{DPS,StopDPS}. An experiment realization of this protocol in a laser-cooled ensemble of a cloud of Rb-87 atoms in a magneto-optical trap achieved a favourable result of 0.036 storage efficiency for a transport distance of 1.26 mm (1/e size of the cloud) for a lifetime of 2.6 ms \cite{transp}. The approach and results show a high promise for EIT in developing robust mechanisms for quantum optical storage with increased transport distance, though reducing decoherence due to natural lifetime and dephasing is still a problem of study. EIT also has great promise in few photon systems and creating schemes for resonant non-linear interactions that create large single-photon phase shifts, which is of great interest in quantum information, such as in the development of high fidelity quantum gates. There have been a number of theoretical and experimental approaches to achieve high fidelity whilst reducing the characteristic residual absorption of EIT \cite{slpg,cqpg}.

Further uses of EIT include transmission with negligible dissipation, non-linear optics in the weak field regime, and slow light. Moreover EIT does not require the highly controlled experimental setups (e.g ultra-low temperatures) for these phenomena to occur. Whilst EIT was initially discovered in atomic/molecular systems (chief of which is the three-level $\Lambda$ system), it has been studied in optomechanical systems, plasmonics, coupled microresonators, solid-state physics and photonic crystals\cite{Whisp}. Especially considering the rise of interest in  quantum metrology using atomic systems, EIT is a powerful candidate to investigate in the increased performance and precision in atomic clocks, in high-resolution atomic interferometry and in magnetoptic measurements. Traditional weaknesses of EIT including low signal-to-noise ratio and fidelity, have been compensated for by the great versatility in control protocols that can be developed to target phenomena of interest. For example, an investigation into the frequency stability of atomic clocks found higher stability when tuning EIT towards detecting magneto-optically induced light polarization shifts rather than controlling the intensity of EIT fields \cite{guidry}. This has naturally led to a great interest in EIT for quantum control theory for developing efficient practical protocols for optimizing performance and precision.

Our study will be on the three-level $\Lambda$ system, a simple starting point to gain a strong understanding in the features and challenges with EIT.

\section{Developing a picture of EIT with three-level $\Lambda$ systems}
We consider the 3-level $\Lambda$ system with stable energy levels $\ket{1}$, $\ket{2}$ and metastable excited state $\ket{3}$. Only the 1-3,2-3 transitions are allowed. We will be considering the dynamics with the 1st Born approximation and the Markovian approximation. Most of our work will be based on a Lindbladian master equation. 

An archetypal example of such a system is for the Rubidium $D_2$ transition fine splittings of the $5S_{1/2}$,$F=1$ and $5P_{1/2}$,$F=2$ energy levels. We take our ground state, $\ket{1}$, as $[5S_{1/2},F=1,m_f=0]$, the excited state, $\ket{3}$, as $[5P_{1/2},F=1,m_f=1]$ and the Raman ground state,$\ket{2}$, as $[5S_{1/2},F=2,m_f=0]$. The 1-3 transition frequency is 377 THz (not including the contributions from the fine splitting differences) and the Stokes shift is 6.384 GHz. The natural lifetime of the transition $\tau=2\pi/\Gamma$=27.7 ns. The hyperfine transition dipole matrix elements for our chosen transitions is 2.54 Debye\cite{ldata}. 

 We first develop our model's interaction Hamiltonian.

\begin{figure}
	\includegraphics[width=\linewidth]{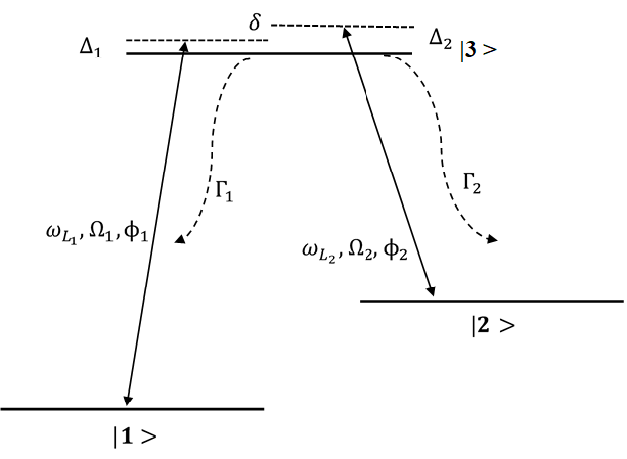}
	\caption{Three level $\Lambda$ system with laser frequency $\omega_{L}$, Rabi frequency $\Omega$, laser phase modulation $\phi$, one-photon detuning $\Delta$, two-photon detuning $\delta=\Delta_{2}-\Delta_{1}$ and decay rate $\Gamma$. }
	\label{fig:3lvl}
\end{figure}

\newpage

\subsection{Hamiltonian in the Field Interaction Picture}
We consider $\ket{3}$ to be the zero energy reference and we define the pump $E_{1}$ and Stokes $E_{2}$ electric fields as below

\begin{equation*}
E_{j}(t) = E^{0}_{j}(t)\cos(\omega_{L_{j}}(t-t_{c})+\phi_{j}(t))
\end{equation*}

The Rabi frequency is given by:
\begin{equation*}
\Omega_{j}(t) = \frac{|p_{j3}|E^{0}_{j}(t)}{\hbar}
\end{equation*}

The dipole moment can be complex, but we introduce the phase function into the oscillatory part of the field. We have:
\begin{align*}
p_{j3} &= |p_{j3}|e^{i\theta_{j}}\\
&\tilde{\phi}_{j}(t) = \phi_{j}(t)+\theta_{j}
\end{align*}

We now define the interaction Hamiltonian $H_{int}$ where:

\begin{equation*}
H_{int}=\hat{{U}}H_{int}\hat{{U}}^\dagger
\end{equation*}

We use the unitary operator to go to the interaction picture and apply the rotating wave approximation (RWA) where:

\begin{equation*}
H_{int}^{j3}=\dfrac{\hbar}{2}\Omega_{1}(t)(e^{i((2\omega_{L_{j}}+\Delta_{j})(t-t_{c})+\tilde{\phi}_{j}(t)}+e^{-i(\Delta_{j}(t-t_{c})+\tilde{\phi}_{j}(t)})
\end{equation*}

\begin{equation*}
H_{int}^{3j}=(H_{int}^{j3})^{*}
\end{equation*}

And all other terms are zero. We ignore the fast oscillating term and transform back to the Schrodinger picture, and now consider the field interaction picture using the unitary operator:

\begin{equation*}
\hat{{U}}(t)=e^{-i\omega_{L_{j}}t/\hbar}
\end{equation*}
\begin{equation*}
\hat{{U}}(t)=e^{-iH_{0}t/\hbar}
\end{equation*}

We arrive at the field interaction Hamiltonian

\begin{equation*}
H_{FE} = \dfrac{\hbar}{2}\begin{pmatrix}
2\Delta_{1} & 0 & \Omega_{1}(t)e^{i\tilde{\phi}_{1}(t)} \\
0 & 2\Delta_{2}  & \Omega_{2}(t)e^{i\tilde{\phi}_{2}(t)} \\
\Omega_{1}(t)e^{-i\tilde{\phi}_{1}(t)} & \Omega_{2}(t)e^{-i\tilde{\phi}_{2}(t)} & 0 
\end{pmatrix}
\end{equation*}.

\subsection{Lindblad Formalism: Finding EIT steady states}
Now we consider the case for open dynamics. We use $\Delta_{2}=\Delta_{1}-\delta$ and define the master equation with the superoperator $\mathcal{L}$:

\begin{equation*}
\dot{\rho}=-\dfrac{i}{\hbar}\comm{H_{FE}}{\rho}+\mathcal{L_\mathcal{J}}\rho
\end{equation*}

\begin{align*}
&\mathcal{L_\mathcal{J}}\rho=\sum_{j\in\mathcal{J}}\lambda_j\mathcal{F}_j(\rho)\\
&\mathcal{F}_j(\rho)=F_j\rho F_j^{\dagger}-\dfrac{1}{2}\left\{F_j^{\dagger}{F}_j,\rho\right\}
\end{align*}

\begin{align*}
&\{F_j\}=\{\ket{1}\bra{3},\ket{2}\bra{3},\ket{1}\bra{1},\ket{2}\bra{2},\ket{3}\bra{3}\}\\
&\{\lambda_j\}=\left\{\dfrac{\Gamma_{1}}{2},\dfrac{\Gamma_{2}}{2},\dfrac{\gamma_{12}+\gamma_{2}-\gamma_{3}}{2},\dfrac{\gamma_{12}-\gamma_{2}+\gamma_{3}}{2},\dfrac{-\gamma_{12}+\gamma_{2}+\gamma_{3}}{2}\right\}
\end{align*}

\begin{equation*}
\mathcal{L}\rho = \begin{pmatrix}
\Gamma_{1}\rho_{33} & -\gamma_{12}\rho_{12} & -(\dfrac{\Gamma_{1}+\Gamma_{2}}{2}+\gamma_{1})\rho_{13} \\
-\gamma_{12}\rho_{21} & \Gamma_{2}\rho_{33}  & -(\dfrac{\Gamma_{1}+\Gamma_{2}}{2}+\gamma_{2})\rho_{23} \\
-(\dfrac{\Gamma_{1}+\Gamma_{2}}{2}+\gamma_{1})\rho_{31} & -(\dfrac{\Gamma_{1}+\Gamma_{2}}{2}+\gamma_{2})\rho_{32} & -(\Gamma_{1}+\Gamma_{2})\rho_{33} 
\end{pmatrix}
\end{equation*}.

We define $\Gamma_{j}$ to be the natural decay rate from the excited state to the jth state and $\gamma$ are dephasing parameters to account for thermal broadening and other losses of coherence.
We then define the dephasing parameters:

\begin{align*}
\gamma_{31} &= \dfrac{\Gamma_{1}+\Gamma_{2}}{2}+\gamma_{1} \\
\gamma_{32} &= \dfrac{\Gamma_{1}+\Gamma_{2}}{2}+\gamma_{2}\\
\gamma &= \gamma_{12}.
\end{align*}

We find stationary solutions for this open system that corresponds to an EIT state, we use the Gell-Mann matrices as our basis and solve for the Bloch vector:

\begin{align*}
\expval{\vec{\sigma}} = \Tr{\vec{\rho\sigma}} \\
\expval{\vec{\sigma}} = (\expval{\sigma_{1}},...,\expval{\sigma_{9}})
\end{align*}

We solve for the density matrix terms in the Bloch equations using the below initial condition:
\begin{align*}
\expval{\dot{\vec{\sigma}}(0)} = 0
\end{align*}

We find the coherence terms, reducing equations to depend on one unknown $\rho_{12}$, assuming the populations for determined parameters are known and fixed: 

\begin{align*}
\rho_{13} &= \dfrac{i\gamma_{31}+\Delta_{1}}{2(\gamma_{31}^{2}+\Delta_{1}^{2})}\left(\Omega_{1}(t)e^{i\tilde{\phi}_{1}(t)}(\rho_{11}-\rho_{33})+\Omega_{2}(t)e^{i\tilde{\phi}_{2}(t)}\rho_{12})\right) \\
\rho_{23} &= \dfrac{i\gamma_{32}+\Delta_{2}}{2(\gamma_{32}^{2}+\Delta_{2}^{2})}\left(\Omega_{1}(t)e^{i\tilde{\phi}_{1}(t)}\rho_{21}+\Omega_{2}(t)e^{i\tilde{\phi}_{2}(t)}(\rho_{22}-\rho_{33})\right).
\end{align*}

The solution to the differential equation for $\rho_{12}$ are obtained by substituting the above expressions for $\rho_{13}$, $\rho_{23}$:

\begin{align*}
\dot{\rho}_{12} &= (-\Gamma'+i\omega')\rho_{12}+K. \\
\Gamma' &= \gamma+\dfrac{\gamma_{31}}{4}\left(\dfrac{\gamma_{31}}{\gamma_{32}}\dfrac{\left(\dfrac{\Omega_{1}}{\gamma_{31}}\right)^{2}}{1+\left(\dfrac{\Delta_{2}}{\gamma_{31}}\right)^2\left(\dfrac{\gamma_{31}}{\gamma_{32}}\right)^{2}}+\dfrac{\left(\dfrac{\Omega_{2}}{\gamma_{31}}\right)^{2}}{1+\left(\dfrac{\Delta_{1}}{\gamma_{31}}\right)^2}\right) \\
\omega' &=
\delta+\dfrac{\gamma_{31}}{4}\left(-\dfrac{\Delta_{2}}{\gamma_{31}}\left(\dfrac{\gamma_{31}}{\gamma_{32}}\right)^{2}\dfrac{\left(\dfrac{\Omega_{1}}{\gamma_{31}}\right)^{2}}{1+\left(\dfrac{\Delta_{2}}{\gamma_{31}}\right)^2\left(\dfrac{\gamma_{31}}{\gamma_{32}}\right)^{2}}+\dfrac{\Delta_{1}}{\gamma_{31}}\dfrac{\left(\dfrac{\Omega_{2}}{\gamma_{31}}\right)^{2}}{1+\left(\dfrac{\Delta_{1}}{\gamma_{31}}\right)^2}\right).
\end{align*}
\begin{multline*}
K=\dfrac{\gamma_{31}}{4}\left(-\left(\dfrac{\gamma_{31}}{\gamma_{32}}\right)^{2}\dfrac{\Omega_{1}\Omega_{2}}{\gamma_{31}^{2}}\dfrac{(\rho_{11}-\rho_{33})e^{i(\tilde{\phi}_{1}(t)-\tilde{\phi}_{2}(t))}}{1+\left(\dfrac{\Delta_{2}}{\gamma_{31}}\right)^2\left(\dfrac{\gamma_{31}}{\gamma_{32}}\right)^{2}}\left(1+i\dfrac{\Delta_{2}}{\gamma_{31}}\dfrac{\gamma_{31}}{\gamma_{32}}\right)\right)\\
+\dfrac{\gamma_{31}}{4}\left(\dfrac{\Omega_{1}\Omega_{2}}{\gamma_{31}^{2}}\dfrac{(\rho_{22}-\rho_{33})e^{i(\tilde{\phi}_{1}(t)-\tilde{\phi}_{2}(t))}}{1+\left(\dfrac{\Delta_{1}}{\gamma_{31}}\right)^2}\left(-1+i\dfrac{\Delta_{1}}{\gamma_{31}}\right)\right)
\end{multline*}

In the weak field approximation, we generally see a significant oscillatory behaviour only for the case of non-zero detuning. The scattering parameter also depends mainly on dephasing between the two ground states when $\gamma_{3j}$ is large. Now we solve for the initial value of $\rho_{12}$ at the stationary point:

\begin{align*}
\rho_{12}^s &= \dfrac{K}{\Gamma'-i\omega'}.
\end{align*}

Next we find the expressions for the steady-state coherences. The expression of $\rho_{13}^s$ and $\rho_{23}^s$ is our main interest as it is a measure of the dispersion and the absorption of radiation near the resonance of the ($\ket{1}\rightarrow\ket{3}$) transition and will be used to determine the optical response functions

\begin{align*}
\rho_{13}^s= \dfrac{\dfrac{\Delta_{1}}{\gamma_{31}}+i}{1+\left(\dfrac{\Delta_{1}}{\gamma_{31}}\right)^{2}}\left(\dfrac{\Omega_{1}}{\gamma_{31}}e^{i\tilde{\phi}_{1}(t)}(\rho_{11}-\rho_{33})+\dfrac{\dfrac{\Omega_{2}}{\gamma_{31}}\dfrac{K}{\gamma_{31}}\left(\dfrac{\Gamma'}{\gamma_{31}}+i\dfrac{\omega'}{\gamma_{31}}\right)e^{i\tilde{\phi}_{2}(t)}}{\left(\dfrac{\Gamma'}{\gamma_{31}}\right)^2 +\left(\dfrac{\omega'}{\gamma_{31}}\right)^2}\right)
\end{align*}

\begin{align*}
\rho_{32}^s= \dfrac{\dfrac{\Delta_{2}}{\gamma_{32}}+i}{1+\left(\dfrac{\Delta_{2}}{\gamma_{32}}\right)^{2}}\left(\dfrac{\Omega_{2}}{\gamma_{32}}e^{-i\tilde{\phi}_{2}(t)}(\rho_{22}-\rho_{33})-\dfrac{\dfrac{\Omega_{1}}{\gamma_{32}}\dfrac{K}{\gamma_{32}}\left(\dfrac{\Gamma'}{\gamma_{32}}+i\dfrac{\omega'}{\gamma_{32}}\right)e^{-i\tilde{\phi}_{1}(t)}}{\left(\dfrac{\Gamma'}{\gamma_{32}}\right)^2 +\left(\dfrac{\omega'}{\gamma_{32}}\right)^2}\right)/
\end{align*}

We note that the system has a natural timescale dependent on the dephasing parameters $\gamma_{3j}$. For the rest of this chapter, we will set $\gamma_{31}=1$.

The above is the general results for the coherences for the steady-state solution. In the weak-field approximation, the fields being smaller than the natural decay rate implies that the steady state is unique as all the other eigenvalues of the Lindbladian have negative real parts. With a model of the microscopic dynamics understood, we now build the optical response functions.

\subsection{Optical response functions}
We find the expectation for the dipole density operator. First, we rotate from the from the corotating field interaction picture back to the Schr\"{o}dinger picture using (9):

\begin{align*}
\tilde{\rho}(t)=U(t)^{\dagger}\rho(t) U(t).
\end{align*}

When we rotate back to the 
ground reference frame, the coherence terms oscillate with the frequencies of the fields driving the transitions:
\begin{align*}
\tilde{\rho}_{3j}(t)=\rho_{3j}(t)e^{-i\omega_{L_{j}}t}.
\end{align*}

We compare the terms for the polarization field in terms of the electric field and susceptibility tensor, and the density matrix description. Here $\chi(\omega,E)$ is a generally non-linear function and depends on the frequency and the amplitude of the electric field. Then the polarization reads:

\begin{align*}
\mathbf{P(\mathbf{r},\mathnormal{t})} &= \chi(t)* E(\mathbf{r},t)= \int_{-\infty}^{t}\chi(t-t')E(\mathbf{r},t')dt'
\end{align*}

If we use a control protocol where we vary the field properties over time or if the envelope is time dependent, it's not obvious what the form of the polarization will be.

In the case where we have a superposition of monochromatic plane waves, the expression simplifies to:
\begin{align*}
\mathbf{P(\mathbf{r},\mathnormal{t})} &= \sum_{k=1}^{n}(E_{0}/2(\chi_{k}(\omega,t)e^{-i\omega_{k}t}+\chi_{k}^{\dagger}(\omega,t) e^{i\omega_{k}t}))
\end{align*}

For the case where we have a general field of the form $E(\mathbf{r},t)=E_{0}(\mathbf{r},t)cos(\omega_{l} t+\phi)$, where we take $\phi$ to be constant, we get the following:

\begin{align*}
\mathbf{P(\mathbf{r},\mathnormal{t})} &= (\chi(\omega,t)* E_{0}(\mathbf{r},\omega))e^{-i\omega_{l}t}+(\chi^{\dagger}(\omega,t)* E_{0}(\mathbf{r},\omega))e^{i\omega_{l}t}.
\end{align*}

For the rest of the chapter, we deal with CW light and monochromatic plane waves.

\begin{multline}
P = \epsilon_{0}\chi(\omega,E)E = \dfrac{\epsilon_{0}\hbar\Omega_{1}}{2|p_{13}|}(\chi^{*}_{1}(\omega,E)e^{i(\omega_{L_{1}}t+\phi_{1})}+\chi_{1}e^{-i(\omega_{L_{1}}t+\phi_{1})}) \\
+\dfrac{\epsilon_{0}\hbar\Omega_{2}}{2|p_{23}|}(\chi^{*}_{2}(\omega,E)(e^{i(\omega_{L_{2}}t+\phi_{2})}+\chi_{2}e^{-i(\omega_{L_{2}}t+\phi_{2})})
\end{multline}

\begin{multline}
\expval{p} = \Tr{\rho p} = \zeta_{1}(\omega,t)+\zeta_{2}(\omega,t) \equiv |p_{13}|/2(\rho_{13}e^{i(\omega_{L_{1}}t+\phi_{1}+\theta_{1})}+\rho_{31}e^{-i(\omega_{L_{1}}t+\phi_{1}+\theta_{1})}) \\
+|p_{23}|/2(\rho_{23}e^{i(\omega_{L_{2}}t+\phi_{2}+\theta_{2})}+\rho_{32}e^{-i(\omega_{L_{2}}t+\phi_{2}+\theta_{2})}).
\end{multline}

We have n as the number density of atoms/molecules:
\begin{align*}
\expval{P} = n\expval{p}.
\end{align*}

We use a simple model of a homogeneous medium of a dilute weakly interacting gas.
Comparing (1) and (2), we see that we have to match oscillating factors to get the electric susceptibility functions. Then it follows that:

\begin{align*}
\chi_{j}(\omega,E) = \dfrac{2n}{\hbar\epsilon_{0}}\dfrac{|p_{3j}|^{2}}{\Omega_{j}}\rho_{3j}e^{-i\theta_{j}}.
\end{align*}

At this point, we have enough information to find other optical response functions such as the complex refractive index, the phase and the group velocities and the Group Velocity Dispersion (GVD). Now we focus our attention in the region in the parameter space in which we develop our model for the EIT optical response functions.

\section{Control scheme}
With the above results, now we can develop a scheme where we explore the slow light phenomenon and the strong dispersive effects of EIT. Our goal is to develop a control scheme for a given atomic system, which yields high transmission of the probe pulses with the group profiles centered near frequencies which is offset to $\omega_{31}$ by a chosen value, therefore belonging to the resonance window. The parameters of the system+field are chosen to satisfy this condition as well as to optimise changes in dispersion properties, such as the group velocity, GVD, and the chirp spectrum, so that small differences in the frequency spectrum of the probe pulses lead to significant deviations in group profiles of pulses with differing spectra propagating in the atomic medium.

\subsection{The concept}
We choose a control Rabi frequency relatively large compared to the pump Rabi frequency which gives a peak near resonance with sharp dispersion and low absorptivity, and we choose a value of the two-photon detuning to shift the transparency window relative to the resonance. Choosing the non-zero detunings does contribute to some population of the other states, though our conditions are still valid for small detunings.
For the rest of the parameters that give us a spike  in the susceptibility near resonance. We use units of $\gamma_{31}$ to make relevant parameters dimensionless (time is in $(\gamma_{31}/2\pi)^{-1}$). We choose the following conditions:

\begin{align}
\begin{split}
\gamma_{31}=1\\
\Omega_{j},\Delta_{j}\ll 1\\
\Omega_{1}\ll\Omega_{2}\\
\rho_{11}\approx 1\\
\omega'\approx\Delta_{2}\nu_{2}-\Delta_{1}\nu_{1}\\
\alpha=\gamma_{32}/\gamma_{31}.
\end{split}
\end{align}

Our next step is to refine the choice of parameters to optimise the optical response of the system to a weak probe pulse of form $E(t)=A(t)e^{i(\omega_{L_{1}}+x)t}$, ($\mu\abs{E(t)}\ll\hbar\Omega_{1}$). The polarization $P(t)$ gains a chirp in the atomic medium and its chirp spectrum is given by $P_{chirp}(\omega)=A_{\omega}(\omega-(\omega_{L_{1}}+x))\chi(\omega)$.  We can define the chirp more explicitly:

\begin{align*}
P_{chirp}(\omega)=\abs{A_{\omega}(\omega-(\omega_{L_{1}}+x))}\abs{\chi(\omega)}e^{i(\phi_{A}(\omega-(\omega_{L_{1}}+x))+\phi_{\chi}(\omega))}.
\end{align*}

Each frequency component gains a phase delay and a group delay, $\tau_p(\omega)$, $\tau_g(\omega)$:

\begin{align*}
&\tau_p(\omega)=-\dfrac{\phi_{\chi}(\omega)}{\omega}\\
&\tau_g(\omega)=-\dfrac{d\phi_{\chi}}{d\omega}.
\end{align*}

The above delays allow us to construct the chirp spectra for probe pulses that can yield phenomena such as anomalous dispersion, group velocity direction changes and pulse lengthening. We consider control of group velocity, GVD and the chirp spectrum in the general case and for transform-limited pulses.

\subsubsection{Control using group velocity, group velocity dispersion and the chirp spectrum}
If we wish to be in the neighbourhood of the minimum group velocities, we choose parameters such that $\left.\dfrac{d^2k'}{d\omega^2}\right|_{\omega=\omega_{L_{1}}+x}=0$ (the local extremum of $\dfrac{dk'}{d\omega}$ in the resonance window). Here, the group velocity differences are small but we can focus on controlling of the dispersion and extinction coefficients at low velocities.
\newline

To maximize the delay between the probe and pump pulses, we intend to maximize $\Delta T\propto \left(\dfrac{1}{v_g(\omega_{L1}+x)}-\dfrac{1}{v_g(\omega_{L1})}\right)$. As a  general approach, we choose parameters where the group velocities are small in the interval $[\omega_{L1},\omega_{L1}+x]$ but the difference in $v_g$ is still significant.

If the second-order Taylor series for $n'(\omega)$ is a good approximation in the interval, the GVD can be used to estimate the time delay/chirp. We choose parameters such that $\dfrac{dn'}{d\omega}$ resembles a linear function near our chosen frequency, say $\omega_{L1}$, giving us local extrema for the GVD. We have $\Delta T\propto GVD(\omega_{L1})x$
\newline

For a probe pulse with a Gaussian envelope, the duration of the pulse (related to the standard deviation $\sigma$) is extended after traveling a distance $d$ in the medium. We define $\sigma_{ex}=\sqrt{\sigma^2+(0.5d\cross GVD(\omega)/\sigma)^2}$ as the extended pulse standard deviation.
\newline

In general, we can always design our optical response function to chirp the spectrum of probe pulses to our purposes. Whilst not discussed in this work, the use of chirped pulses for the driving and probe fields can subdue undesired non-adiabatic effects during pulse propagation in a medium and add more control to shaping of the optical response functions and the chirp spectra of post-medium pulses\cite{Klaas}. For sustaining EIT states in general high dimensional systems, spectral modulation by methods such as linear and sinuosidal spectral chirp can improve mitigation of decoherence effects and adiabatically transfer populations\cite{Rydmbs, Harmspec}. 
\newpage

In the next section, we develop approximate forms for the optical response functions and its derivatives in the control regime.

\subsection{Optical response functions in the control regime}
We consider the approximate form of the susceptibility $\chi=\chi'+i\chi''$ near resonance. We use $\Delta\omega=\omega-\omega_{31}$ as the argument of the functions.

\begin{align}
	\begin{split}
		\chi'(\Delta\omega)\approx\dfrac{2n|p_{31}|^{2}}{\hbar\epsilon_{0}}&\left((\rho_{11}-\rho_{33})\Delta\omega+\dfrac{Q_1\Gamma'(\Delta_{2}\nu_{2}-\Delta\omega\nu_{1})}{(\Gamma'^2+(\Delta_{2}\nu_{2}-\Delta\omega\nu_{1})^{2})}\right.\\
		&\left.\text{ }+\dfrac{Q_2\Gamma'\Delta_{2}}{(\Gamma'^2+(\Delta_{2}\nu_{2}-\Delta\omega\nu_{1})^{2})}\right)
	\end{split}
\end{align}
\begin{align}
	\chi''(\Delta\omega)\approx\dfrac{2n|p_{3j}|^{2}}{\hbar\epsilon_{0}}&\left((\rho_{11}-\rho_{33})-\dfrac{Q_3\Gamma'}{(\Gamma'^2+(\Delta_{2}\nu_{2}-\Delta\omega\nu_{1})^{2})}\right)
\end{align}
\begin{align*}
	&Q_1=\dfrac{\Omega_2^2}{4\Gamma'}(\alpha^2(\rho_{11}-\rho_{33})+(\rho_{22}-\rho_{33}))\\
	&\text{ }\text{ }\text{ }\text{ }+\dfrac{\alpha}{\nu_1}(\rho_{11}-\rho_{33})+(\rho_{22}-\rho_{33})\\
	&Q_2=\alpha(\rho_{11}-\rho_{33})-\dfrac{\alpha}{\nu_1}(\rho_{11}-\rho_{33})+(\rho_{22}-\rho_{33})\\
	&Q_3=\dfrac{\Omega_2^2}{4}(\alpha^2(\rho_{11}-\rho_{33})+(\rho_{22}-\rho_{33})).
\end{align*}

We see that the real susceptibility is the sum of a linear term that depends on the population in the 1st ground state plus a distribution (similar to a Fano resonance) with the maximum value $(Q_1)/(2\Gamma')$ and $2\Gamma'$ being the full width between the local extrema. The second term is an overlapping Cauchy distribution with FWHM $\Gamma'$ and amplitude $Q_2\Delta_2(\Gamma')^{-1}$. Similarly, the imaginary part of the susceptibility has a constant term plus a Cauchy distribution with FWHM $2\Gamma'$.

\begin{figure}
	\includegraphics[width=\linewidth]{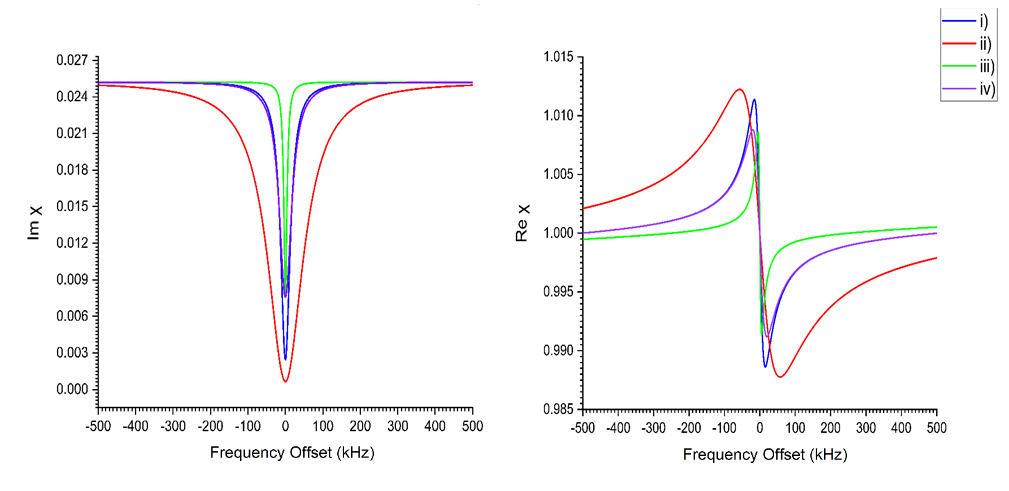}
	\caption{Plot of the imaginary (left) and the real(right) part of the susceptibility of the system for parameters $(\Omega_{2},\gamma_{12})$ for the numbered configurations: i)$(1/35,4.3E-5)$, ii)$(2/35,4.3E-5)$ iii)$(1/70,4.3E-5)$ iv)$(1/30,1.7E-4)$ and $\Omega_1=0.1\Omega_2$.}  
	\label{fig:4lvl}
\end{figure}

Using our scheme, we assume the following approximations for the parameters:

\begin{align*}
\Gamma'\approx\gamma+\gamma_{31}\left(\dfrac{\Omega_{2}^{2}}{4}+\dfrac{\alpha\Omega_{1}^{2}}{4}\right)\\
\nu_1=(1-\gamma_{31}\Omega_{2}^{2}/4)>0 && \nu_2=(1-\gamma_{31}(\alpha\Omega_{1})^{2}/4)>0.
\end{align*}

Below, we use the approximations for the dispersion and extinction coefficients:

\begin{align*}
n'(\omega)\approx 1+0.5\chi'(\omega) && n''(\omega)\approx\dfrac{\chi''(\omega)}{2}.
\end{align*}

\begin{figure}
	\includegraphics[width=\linewidth]{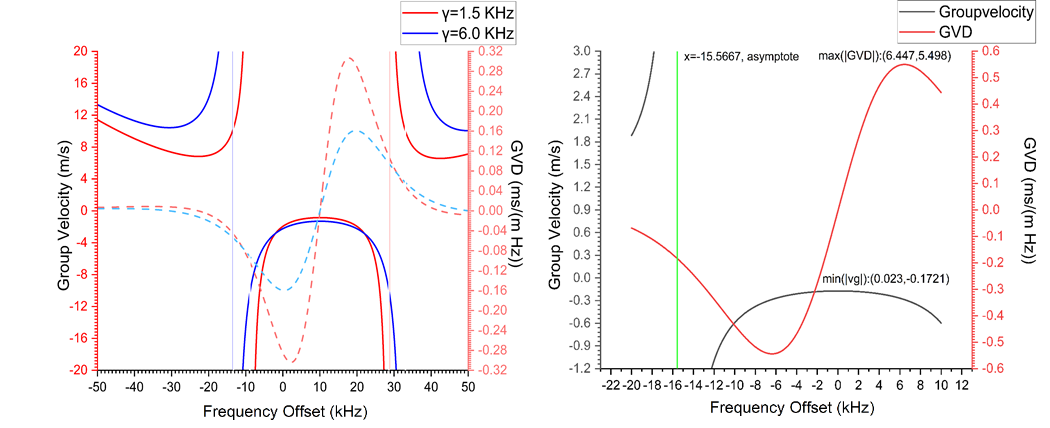}
	\caption{(Left) Plot of the group velocity (solid line) and the GVD (dashed line) for the configuration in Sec 2 for two values of $\gamma_{12}$ 1.5 kHZ, 6.0 kHZ and $\Omega_2=1/35$, $\Omega_{1}=0.1\Omega_{2}$. (Right) Plot of group velocity and the GVD showing anomalous dispersion and local minima.}  
	\label{fig:2lvl}
\end{figure}

\begin{figure}
	\includegraphics[width=\linewidth]{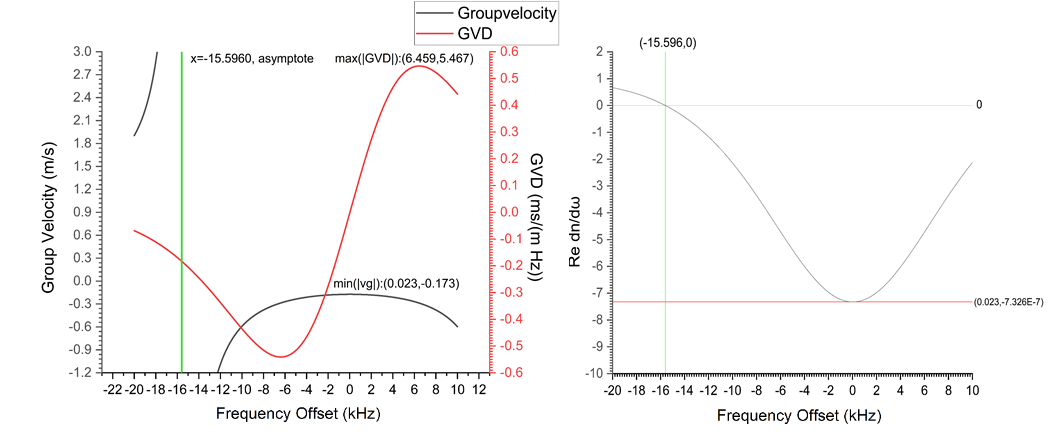}
	\caption{(Left) Plot of the group velocity and the GVD showing anomalous dispersion and local minima with $\Omega_{1}$ scaled by factor of 1.1x. Green line is the asymptote where anomalous dispersion occurs and the group velocity becomes infinite. We see how small perturbations in the pump Rabi frequency cause non-trivial changes in the dispersion of a pulse propagating through the medium. (Right) Plot showing the first derivative of the dispersion and the zero and the local minima/maxima in the derivative correspond respectively to anomalous dispersion asymptotes and the local maxima/minima of the dispersion.}  
	\label{fig:2lvl}
\end{figure}

We get the expression for the group velocity. For brevity, we define ,$f(\Delta\omega)=(\Gamma')^2+(\Delta_{2}\nu_{2}-\Delta_1\nu_1)^2)$, then:

\begin{align}
	\begin{split}
		v_g(\Delta\omega)/c&\approx\dfrac{(f(\Delta\omega))^2}{\chi_0 Q_1\Gamma'\nu_1}\\
		&\cross\left(\dfrac{(f(\Delta\omega))^2}{\chi_0 Q_1\Gamma'\nu_1}(1+0.5\chi_0(\rho_{11}-\rho_{33})\Delta\omega)\right.\\
		&\left.\text{ }\text{ }\text{ }+\dfrac{f(\Delta\omega)}{2 Q_1\Gamma'\nu_1}(Q_1\Gamma'(\Delta_{2}\nu_{2}-\Delta\omega\nu_1)+Q_2\Gamma'\Delta_2)\right.\\
		&\left.\text{ }\text{ }\text{ }+0.5(\omega_{31}+\Delta\omega)\left(\dfrac{(f(\Delta\omega))^2}{ Q_1\Gamma'\nu_1}(\rho_{11}-\rho_{33})-f(\Delta\omega)+\dfrac{2}{Q_1\Gamma'}(\Delta_{2}\nu_{2}-\Delta\omega\nu_1)\right)\right)^{-1}.
	\end{split}
\end{align}

The above expression implies that we can get infinite and negative group velocities if $\chi'$ has a negative gradient region with a large enough steepness.
We note the group velocity about the saddle point near 0 of the real susceptibility reads:
\begin{align*}
\begin{split}
\min_{\abs{v_g}}{v_{g,e}/c}&\approx \beta_{\text{min}}\left(1+\mathcal{O}\left(\dfrac{\Gamma'}{\omega_{31}}\right)\right)^{-1}\\
&=\dfrac{\Gamma'}{2\chi_0 Q_1\nu_1\omega_{31}}\left(1+\dfrac{\Gamma'}{Q_1\nu_1\omega_{31}}\left(2\Gamma'+\Delta_2\left(Q_2+(\rho_{11}-\rho_{33})\Gamma'\dfrac{\nu_2}{\nu_1}\right)\right)\right)^{-1}
\end{split}
\end{align*}

This is pretty close to the minimum (unsigned) group velocity, and we note that the second term $\ll 1$. We see the strong dependence of the slow light condition on $\Gamma'$ and $\omega_{31}$. Whilst we can use weaker fields to reduce the line broadening effect, the dephasing between the ground states (and therefore $\gamma_{31}$) sets a hard limit on the lowest value of $\Gamma'$. 

With this, we can calculate the group velocity dispersion. In our scheme, our light fields are at least of THz frequencies, therefore the dominant terms in the GVD are those which include $\omega$. We also use the fact that for a dilute medium $\chi',\chi''\ll 1$. Then:

\begin{align}
	\begin{split}
		GVD(\Delta\omega)c&\approx -A\nu_1(\Delta_2\nu_2-\Delta\omega\nu_1)f(\Delta\omega)^{-2}\\
		&\cross(1+(\Delta_2\nu_2-\Delta\omega\nu_1)f(\Delta\omega)^{-1})
	\end{split}
\end{align}
\begin{align*}
	A&=2\omega\chi_{0}\dfrac{\Omega_2^2}{4}(\nu_{1}(\alpha^2(\rho_{11}-\rho_{33})+(\rho_{22}-\rho_{33}))\\
	&+\Gamma'(\alpha(\rho_{11}-\rho_{33})+(\rho_{22}-\rho_{33}))).
\end{align*}

For a light pulse, each frequency component $E(\omega)$ picks up a phase $\Delta\phi(\omega)=k(\omega)d$ after it passes through a dispersive planar medium of length $d$. The pulse exiting the medium will be the Inverse Fourier Transform (IFT) of the post-medium $E(\omega)$ and the time difference between peaks of two simultaneous pulses of frequencies $\omega_1,\omega_2$ is given by:

\begin{equation*}
T=d\left(\dfrac{1}{v_g(\omega_1)}-\dfrac{1}{v_g(\omega_2)}\right).
\end{equation*}

Since the dispersion in the transparency window is strong in our scheme, a transform-limited pulse will have its spectral components accumulate different phases with the time-domain pulse broadening. The GVD is part of the estimate ($\text{GVD}(\omega_0)(\text{bandwidth})d$) of the chirp introduced to the pulse (valid for relatively small pulse bandwidths). Therefore, the group velocity, GVD and the chirp spectrum are effective tools to monitor responses to perturbations in the EIT medium by the way of determining how the properties of the probe pulses change relative to the baseline case. 

An example we will consider is the passage of high-energy charged particles (high energy particle) that induce Cherenkov radiation production in the medium. The probability of a photon being emitted by an individual high energy particle event is rather small and unlikely to be detectable. Tuning the medium to have Cherenkov radiation about frequencies which have slow group velocities will effectively ensure the buildup of the radiation over time as more events pass. In the dark polariton picture, the Cherenkov radiation will couple to the atomic transitions and travel as polaritons. The effect will be to introduce a growing electric potential that will affect the EIT state, which will manifest in probe pulses having a different chirp spectrum upon detection. We will explicitly consider the Cherenkov radiation produced in the EIT medium and give an analytic form to the field in our parameter regime.

\section{Application: Detecting high-energy charged particles using EIT}
Traditional detection of high energy charged particles have been handled in experiments using Cherenkov gas counters where ultra-sensitive detectors measure the small yield of light (essentially 1 photon per 10,000 events). The small yield of Cherenkov radiation, which can be calculated by the Frank-Tamm formula, makes the use of such ultra-sensitive detectors a necessity. But there has been a recent surge of interest in using the techniques in atomic interferometry to measure low-energy physical phenomena; and  using EIT with atomic spectroscopic techniques to measure changes in the atomic medium has been proposed in the 2000s \cite{Slight}. 

We develop on these techniques and propose a semiclassical model which gives an approximate analytic solution for the Cherenkov radiation using the optical response functions calculated using the Lindbladian. The strong non-linear dispersion and high transmission properties of EIT can be used to create Cherenkov radiation with slow group velocities for a train of relativistic charged particles. At a classical level, the slow light condition ensures that the group profile of additive Cherenkov contributions will not smear and constructively increase in intensity. This is owing to the the quantum dark-polariton description which implies that the Cherenkov radiation develops coherence with the transition dipoles after emission by the medium, essentially travelling as slow-moving dark polaritons confined to the medium for a duration proportional to the inverse of the group velocity. Here the  model is developed for the case of using a very dilute homogeneous gaseous medium, but the results can easily be extended and optimised towards more complex systems such as EIT diffraction gratings and crystals.

\subsection{A  model using Fourier transformed Maxwell's equations}

We consider a charged high energy particle traveling with relativistic velocity $\mathbf{v}=\beta c\hat{z}$ and having charge q. We use the treatment of a classical point particle, assuming that the width of its wavefunction is much smaller than the wavelengths of emitted radiation.
The below derivations are in the far-field regime where we assume the particle is sufficiently far away to neglect near-field radiation. The current density and its Fourier transform in the space and time domain is given by:
\begin{align*}
\mathbf{J} = \rho(x,y,z,t)\mathbf{v}=q\delta(z-vt)\delta(x)\delta(y)\mathbf{v}\\
\mathbf{J}_{\omega,k} = (2\pi)q\tilde{\delta}(\omega-\mathbf{k}\cdot\mathbf{v})\mathbf{v}.
\end{align*}

(where $\tilde{\delta}(z)$ is the generalized delta function and $q$ is the charge measured in the lab frame $q=\gamma q_0$).

We solve for the fields produced as a result of this current when placed in a homogeneous, anisotropic, non-magnetic medium driven to a EIT steady state. We use the derived optical response function ($\epsilon(\omega)=1+\chi_{31}(\omega-\omega_{31})$) with the Fourier transformed Maxwell equations to calculate the fields in the far-field limit. The effects of retarded time are neglected in this derivation.  We start with the Fourier transformed Maxwell's equations for an anisotropic, non-magnetic medium:

\begin{align*}
\omega\mathbf{k}\cdot(\epsilon_{0}\check{\epsilon}\cdot \mathbf{E_{\omega,k}}) =& \rho_{\omega,\mathbf{k}}\\
i\omega\mathbf{k}\cdot\mathbf{B_{\omega,k}} =& 0 \\
\mathbf{k}\cross\mathbf{E_{\omega,k}} =& -\omega\mathbf{B_{\omega,k}} \\ 
\mathbf{k}\cross\mathbf{B_{\omega,k}} =& -i\mu\cdot\mathbf{J_{\omega,k}}+\omega\mu\epsilon_{0}\check{\epsilon}\cdot\mathbf{E_{\omega,k}}.
\end{align*}

And then introduce the scalar and vector potentials:

\begin{align*}
&\mathbf{E_{\omega,k}} = i\mathbf{k}V_{\omega,\mathbf{k}}+i\omega\mathbf{A_{\omega,k}}\\
&\omega\mathbf{k}\cdot(\check{\epsilon}\cdot\mathbf{A_{\omega,k}})+(\mathbf{k}\cdot(\check{\epsilon}\cdot \mathbf{k}))V_{\omega,\mathbf{k}} = \dfrac{\rho_{\omega,\mathbf{k}}}{\epsilon_0}\\
&\mathbf{k}\cross(\check{\mu}^{-1}\cdot(\mathbf{k}\cross\mathbf{A_{\omega,k}}))+\dfrac{\omega^2}{c^2}\check{\epsilon}\cdot(\mathbf{A_{\omega,k}}+\dfrac{c}{\omega}\mathbf{k}V_{\omega,\mathbf{k}})=-\mu_0\mathbf{J_{\omega,k}}.
\end{align*}

Here $\check{\epsilon}$ is the 3x3 rank 2 dielectric tensor and is a function of frequency. We choose to take our coordinate system to be along the dielectric axes $(e_{+}, e_{-}, e_{z})$ where the tensor is diagonal. We consider the case where only one transverse mode is active and the medium is nearly homogeneous. We use the Lorentz gauge condition

$\mathbf{k}\cdot(\check{\epsilon}\cdot\mathbf{A_{\omega,k}})+\dfrac{\omega\abs{\check{\epsilon}\cdot\check{\mu}\cdot\check{\epsilon}}}{c^2} V_{\omega,\mathbf{k}}=0$

to arrive at the Helmholtz equations for the potentials \cite{Mxwell}:

\begin{align}
&\mathbf{A_{\omega,k}}=(2\pi)\mu_0q\left(k^2 -\dfrac{\omega^2}{c^2}\epsilon\right)^{-1}\tilde{\delta}(\omega-\mathbf{k}\cdot\mathbf{v})\mathbf{v}\\
&V_{\omega,\mathbf{k}}=(2\pi)\dfrac{q}{\epsilon\epsilon_{0}}\left(k^2 -\dfrac{\omega^2}{c^2}\epsilon\right)^{-1}\tilde{\delta}(\omega-\mathbf{k}\cdot\mathbf{v}).
\end{align}

For lossy mediums, $\mathbf{k^*}\cdot\mathbf{E}\not=0$ in general. We will see the presence of both tranverse and longitudinal modes.

We take the inverse spatial Fourier transform of $\mathbf{A}_{\omega,k}$:
\begin{align*}
\mathbf{A}_{\omega}\cdot\hat{v} = \dfrac{1}{(2\pi)^2}\int_{\Gamma_k}\mu_0qv\left(k^{2}-\dfrac{\omega^{2}}{c^{2}}\check{\epsilon}\cdot\right)^{-1}\tilde{\delta}(\omega-\mathbf{k}\cdot\mathbf{v})e^{i\mathbf{k}\cdot\mathbf{r}}d^2k_{\perp}dk_z.
\end{align*}

Expressing it in cylindrical coordinates and carrying out integration over $k_z$, we get:
 
\begin{align*}
	\mathbf{A}_{\omega}\cdot\hat{v} = \mu_0q e^{i\omega(z/v)}\dfrac{1}{(2\pi)^2}\int_{0}^{\infty}\left(k^{2}-\dfrac{\omega^{2}}{c^{2}}\check{\epsilon}\cdot\right)^{-1}\left(\int_{0}^{2\pi}e^{ik_{\perp}r_{\perp}\cos(\theta)}d\theta\right)k_{\perp}dk_{\perp}
\end{align*}

The angular integral is a representation of the zeroth Bessel function.
 
The above integral, for the $k_\perp$ integrand, has 2 poles $\pm ik_r=\pm\omega c^{-1}\sqrt{\beta^{-2}-\epsilon(\omega)}$ in the complex domain corresponding to forward and backward radially propagating modes. Since the medium is dissipative, we consider only the physical forward propagating pole and use a semi-circular contour with bumps excluding the poles. $k_z=\omega v^{-1}$ fixes the possible values for the radial wavevector. For the case of no loss, the Cherenkov condition is $\Re{\epsilon}>\beta^{-2}$. We will use this later to determine the contributions for radiative modes and evanescent modes:

\begin{align*}
	\mathbf{A}_{\omega}\cdot\hat{v} = \mu_0q e^{i\omega(z/v)}\dfrac{1}{2\pi}\int_{0}^{\infty}\dfrac{k_{\perp}J_0(k_{\perp}r_{\perp})}{k_{\perp}^2+\dfrac{\omega^2}{c^2}\left(\beta^{-2}-\epsilon(\omega)\right)} dk_{\perp}.
\end{align*}

The solution of the integral is the modified Bessel function $K_0$ evaluated at $ik_r$. Considering the treatement in \cite{Saff} in extending the lossless case to the lossy case when we consider transforming to the time domain, the frequency integration can be segmented into two regions $\Re(\epsilon(\omega)-\beta^{-2})\gtrless 0$, which correspond respectively to the radiative and non-radiative regions. We define the radiative region: $\Gamma_1= [\omega_{31}+a,\omega_{31}+b]$. We hence solve the final integral for $\mathbf{A_{\omega}}$, and similarly define $V_{\omega}$:

\begin{align}
	\mathbf{A}_{\omega}(\mathbf{r},\omega)\cdot\hat{v } = \begin{cases}
		\dfrac{\mu_0q}{4\pi}e^{i\omega(z/v)}K_0(ik_r r_{\perp}) & \omega\in[\omega_{31}+a,\omega_{31}+b]^c\\
		\text{ }&\text{ }\\
		\dfrac{\mu_0q}{4\pi}e^{i\omega(z/v)}\dfrac{i\pi}{2}H_0^{(1)}(k_r r_{\perp}) & \omega\in[\omega_{31}+a,\omega_{31}+b]\\
	\end{cases} 
\end{align}

\begin{align}
	V_{\omega}(\mathbf{r},\omega) = \begin{cases}
		\dfrac{q}{4\pi\epsilon_{0}\epsilon v}e^{i\omega(z/v)}K_0(ik_r r_{\perp}) & \omega\in[\omega_{31}+a,\omega_{31}+b]^c\\
		\text{ }&\text{ }\\
		\dfrac{q}{4\pi\epsilon_{0}\epsilon v}e^{i\omega(z/v)}\dfrac{i\pi}{2}H_0^{(1)}(k_r r_{\perp}) & \omega\in[\omega_{31}+a,\omega_{31}+b].\\
	\end{cases}
\end{align}

In the far-field limit, $kr_{\perp}\gg 1$, asymptoptic expansion of the Hankel function gives us $i\pi H_0^{(1)}(k_r r_{\perp})=\sqrt{i\pi}e^{ik_r r}/\sqrt{k_r r}$. We can define the steepest descent contours by using the phase term:

\begin{align}
	\begin{split}
		&\dfrac{r_{\perp}}{c}\theta(\tilde{\omega},r,z,t)=\dfrac{r_{\perp}}{c}i\left(\dfrac{\tilde{\omega}c}{r}\left(\dfrac{z}{\tilde{v}}-t\right)+{\tilde{\omega}n'_{\perp}}+i{\tilde{\omega}n''_{\perp}}\right)\\
	\end{split}
\end{align}

The steepest descent contour will keep the imaginary part of the phase constant whilst the real part will have an additional term $-p$. We define the contour $\gamma_{\tilde{\omega}}(p)$, parameterized by real $p$ by:

\begin{align*}
	\theta(\gamma_{\tilde{\omega}}(p),r,z,t)=\theta(\tilde{\omega},r,z,t)-p
\end{align*}

About a point $\omega_i$, if we can define a local analytic expansion of $\theta(\tilde{\omega},r,z,t)$ extended to the complex plane, we can define an inverse function for $\theta$ and get $\gamma_{\tilde{\omega}}(p)=\theta^{-1}(\theta(\omega_i,r,z,t)+ip)$.
\newline

The method of steepest descent can be used to find any stationary points in the case when $\chi''$ is very small. These group modes represent points of constructive interference of wavefronts on a group cone with vertex at $(0,vt)$. This involves finding the critical points of (12). Analyticity of the refractive index ensures that the critical points will be saddle points and there'll be no local extrema in the complex plane. The critical frequencies $\tilde{\omega_i}$ obey the below condition:

\begin{align}
	t-\dfrac{z}{v}-r_{\perp}\dfrac{dk_{r}}{d\tilde{\omega}}=0.
\end{align}

We deform the integration contour on the real interval to the complex plane such that we pass through all critical points via their steepest descent contours and all other contributions are on segments, which are very small. For the time-domain fields in the radiative domain, we define the set of critical points $\mathcal{S}_c$ with each critical point defined on contour $\gamma_{\tilde{\omega}_i}(p)$. A larger upper limit  $P$ will give a better approximation of the integral

\begin{align}
	V(\mathbf{r},t)\approx\sum_{w_i\in \mathcal{S}_c}
		\dfrac{iq}{4\pi\epsilon_0v}e^{ir_{\perp}c^{-1}\theta(\tilde{\omega}_i,r,z,t)}\dfrac{\sqrt{i\pi}}{\sqrt{2r_{\perp}}}\int_{0}^{P}\dfrac{e^{-r_{\perp}c^{-1}p}}{\epsilon\sqrt{k_{\perp}(\gamma_{\tilde{\omega}_i}(p))}}\dfrac{d\gamma_{\tilde{\omega}_i}}{dp}dp.
\end{align}

In the case where the critical points are non-degenerate and represent local maxima of the real part of the phase, we can use Morse's Lemma for complex functions to get a nicer form of the integral. Since the functions are holomorphic, there exists a neighbourhood near non-degenerate saddle points where phase $\theta(\tilde{\omega})$ is approximately quadratic. Accordingly, we can define the asymptotic expansion of (14) 

\begin{align}
	\begin{split}
		V(\mathbf{r},t) &\approx \dfrac{-iq}{4\pi\epsilon_{0}r_{\perp}v}\sum_{\tilde{\omega}_i\in S_{t,z,r_{\perp}}}\dfrac{e^{r_{\perp}c^{-1}i\theta(\tilde{\omega}_i,t,z,r_{\perp})}}{(-\text{GVD}_r(\tilde{\omega}_i))^{1/2}}\left(\dfrac{1}{\sqrt{\text{k}_{\perp}(\tilde{\omega}_i)}}+\mathcal{O}\left(\dfrac{c}{r}\right)\right).
	\end{split}
\end{align}

The critical points will be complex for a dissipative system and will be located near neighbourhoods of critical points for the non-dissipative system if the dissipation is weak. In the case where no critical points exist or the integration is difficult, we can calculate the Gauss-Laguerre quadrature of the integral \cite{GLQ}. However, when the dissipation is significant, the validity of the Cherenkov condition and the simplicity of the fields is not generally true.

\begin{figure}
	\includegraphics[width=\linewidth]{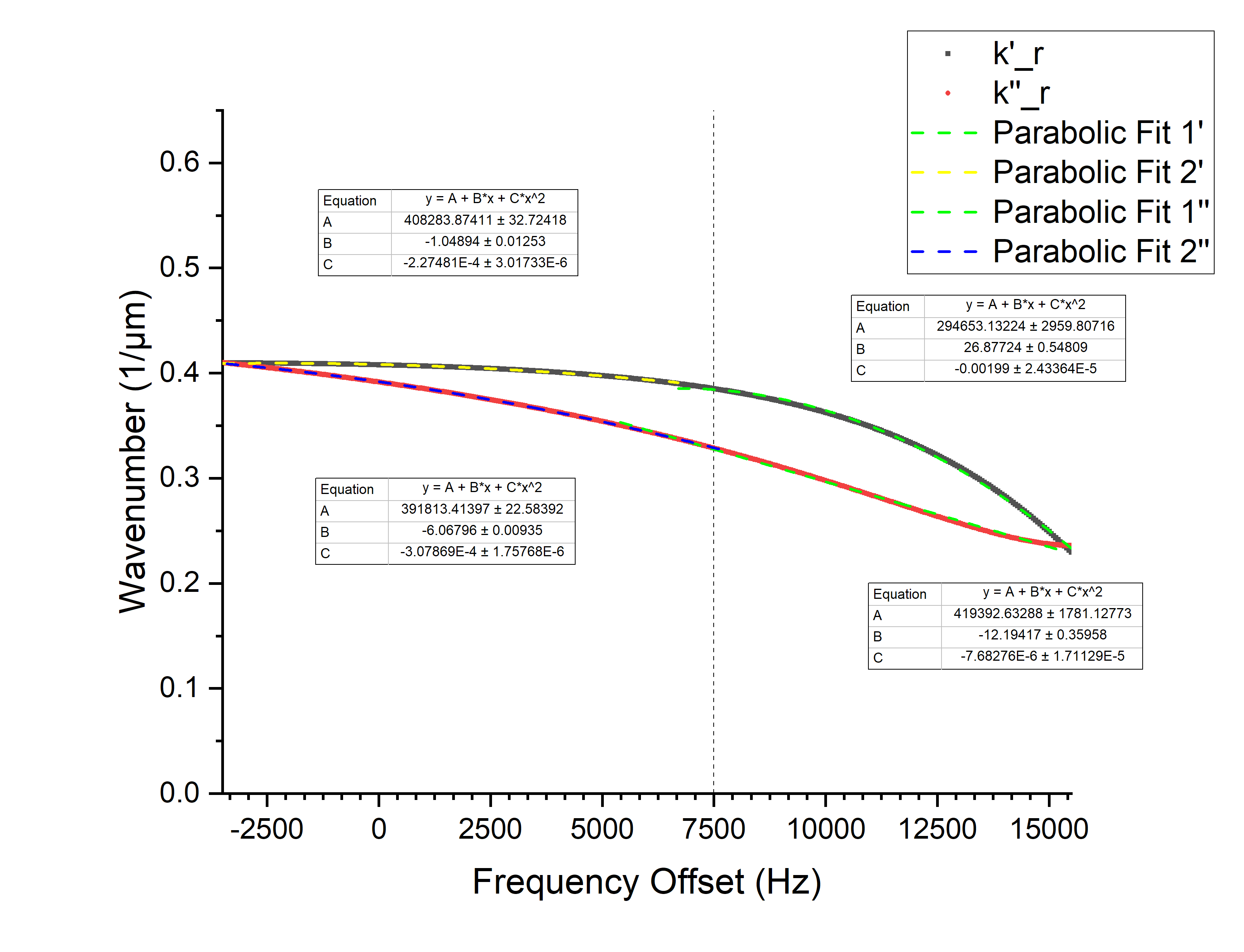}
	\caption{Plots of real and imaginary parts of $k_r$, when the radiative modes are in the anomalous dispersion region, with piecewise parabolic approximations as functions of the frequency offset. For our particular system+field choice, we divide the radiative region into 2 regions with common endpoint at 7500 Hz. Resonant points where $\Re\epsilon=\beta^{-2}$ are at the intersections of real and imaginary parts, bounding the radiative region.}  
	\label{fig:2lvl}
\end{figure}
\newpage

An explicit evaluation of the integral in terms of complex error functions can be done in the far-field limit by approximating $k_{\perp}$ by a piecewise function composed of parabolas. As shown in Figure 5, we consider the local quadratic best fit in the region and expand it about a central frequency, $\omega_{31}'=\omega_{31}+x$, in region $[a_i,b_i]$, $k_r\sim k_0+\eta\Delta\omega+\sigma'(\Delta\omega)^2$. We evaluate the radial component of the E-field which has a Hankel function of first type with order 1. We use the asymptotic far-field limit and we get the below integral:

\begin{align}
	\begin{split}
		\mathbf{E}(\mathbf{r},t) &\approx \dfrac{iq}{8\pi\epsilon_{0}r_{\perp}v}\sqrt{ir_{\perp} \pi} \sum_{j\in I}e^{i\omega'_{31}(z/v-t)+ik_0r_{\perp}}\\
		&\cross\int_{a_j}^{b_j}dw\text{ }e^{i(w(z/v-t)+r_{\perp}(\eta w+\sigma' w^2))}\dfrac{\sqrt{k_r}\hat{r}+\dfrac{\omega}{v\sqrt{k_r}}(1-\beta ^{2}\epsilon)\hat{z}}{\epsilon(w)}
	\end{split} 
\end{align}

\begin{align}
	\begin{split}
		\mathbf{B}(\mathbf{r},t) &\approx \dfrac{iq}{8\pi\epsilon_{0}r_{\perp}c^2}\sqrt{ir_{\perp} \pi} \sum_{j\in I}e^{i\omega'_{31}(z/v-t)+ik_0r_{\perp}}\\
		&\cross\int_{a_j}^{b_j}dw\text{ }e^{i(w(z/v-t)+r_{\perp}(\eta w+\sigma' w^2))}\dfrac{\sqrt{k_r}}{\epsilon(w)}\hat{\phi}.
	\end{split}
\end{align}

We approximate the part of the integrand outside the exponential by a Taylor series in $w$ with coefficients $c_j$. The solution to (16) is a sum of complex error functions. The integral of the exponential term is given below:

\begin{align}
	\begin{split}
	G_0(a,b)=&i\tilde{\sigma}^{-1}\left[(e^{\tilde{\sigma}^2 b^2r_{\perp}-it'b}-e^{\tilde{\sigma}^2 a^2r_{\perp}-it'a})e^{\Phi^2t'^2/(2\sigma^2r_{\perp})}\right.\\
	&\left.-e^{\tilde{\sigma}^2 b^2r_{\perp}-it'b}\text{erfcx}\left(\Phi\left(\dfrac{t'}{\sqrt{2r_{\perp}}\sigma}+i\dfrac{b\tilde{\sigma}^2\sqrt{r_{\perp}}}{\sqrt{2}\sigma}\right)\right)\right.\\
	&\left.+e^{\tilde{\sigma}^2 a^2r_{\perp}-it'a}\text{erfcx}\left(\Phi\left(\dfrac{t'}{\sqrt{2r_{\perp}}\sigma}+i\dfrac{a\tilde{\sigma}^2\sqrt{r_{\perp}}}{\sqrt{2}\sigma}\right)\right)\right]
	\end{split}
\end{align}

\begin{align*}
	\begin{split}	
		&\sigma=\abs{\tilde{\sigma}}\equiv\abs{\sqrt{2i\sigma'}}\text{ }\text{ }\text{ }\text{ }\text{ }\text{ }\text{ }\text{ }\text{ }\Phi=e^{-i\arg(\tilde{\sigma})/2}\text{ }\text{ }\text{ }\text{ }\text{ }\text{ }\text{ }\text{ }\text{ }t'=(t-z/v)-\eta r_{\perp}.
	\end{split}
\end{align*}

The general expansion for any of the fields terms with expressions of form $A\sum_{j\in I}c_j\int_{a_j}^{b_j}dw\text{ }e^{i(w(z/v-t)+r_{\perp}(\eta w+\sigma' w^2))}f(w)$ is given by:

\begin{align}
	F(t,z,r)=A\sum_{j\in I}c_j\sum_{k=0}^{N}i^n\left.\dfrac{d^k f}{dx^k}\right|_{x=\omega'_{31}}\dfrac{d^k}{d(z/v-t)^k}G_0(a_j,b_j).
\end{align}

The expression above provides more information about the relations between the parameters in the Cherenkov field and system parameters. It's clear that $\eta_j$ and $\sigma_j$ are both related to derivatives of the radial wavenumber $k_r$ about the central frequency $x$ for each frequency interval $[a_j,b_j]$. We can estimate these values by looking at the susceptibility and seeing how altering parameters like $\Gamma',\Delta_2,\gamma_{12}$ affect the curvature and first derivative of $k_r$. We see that decreasing the nonlinearity in the susceptibility about region $I_j$ reduces the inhomogeneous broadening of spectral components over propagation distance, suggesting that the linear dispersion region of EIT is ideal for decreasing dissipation and for controlling group velocities.

\begin{figure}
	\includegraphics[width=\linewidth]{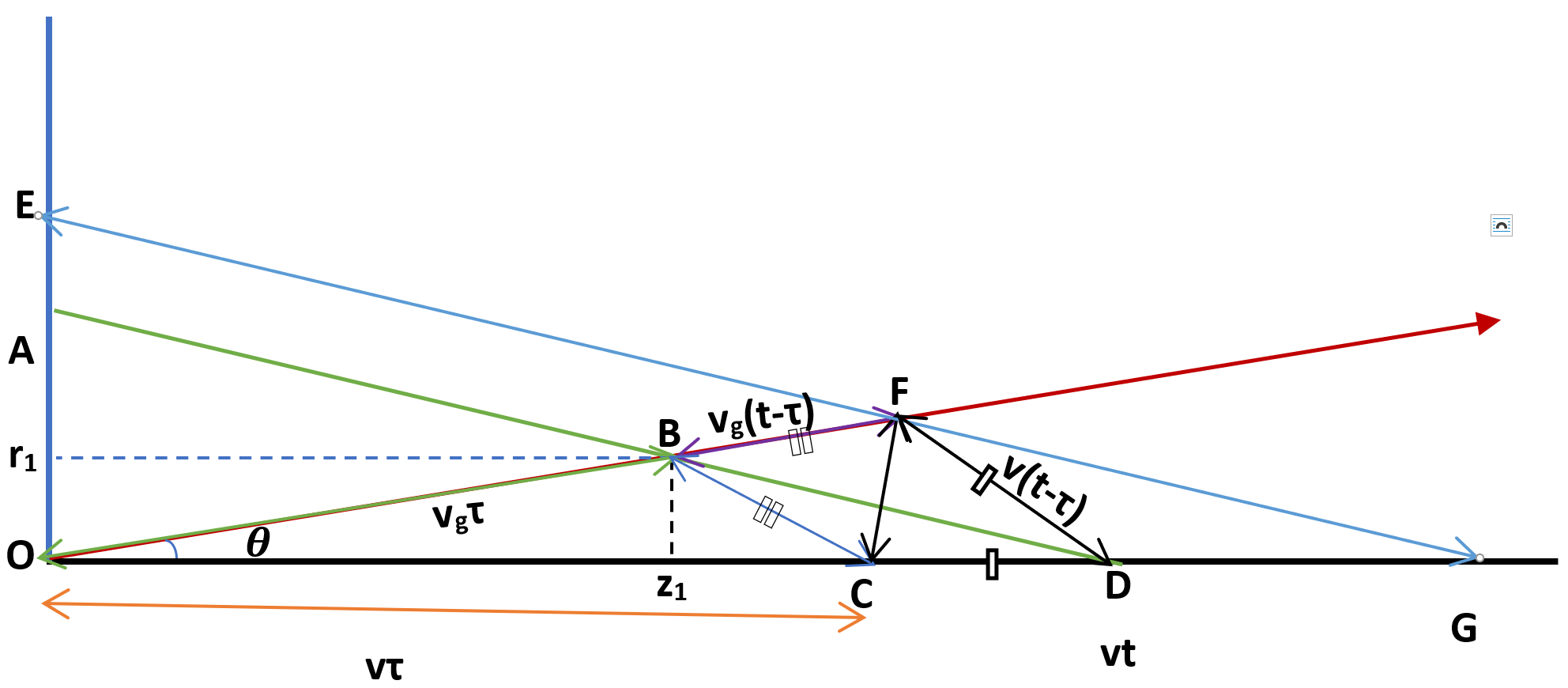}
	\caption{Formation of the group cone for coherent Cherenkov radiation, showing the propogation of the group mode with angle $\theta$ and the radiation propagation distances over transit time $\tau$ and time of Cherenkov cone vortex $t$.}  
	\label{fig:2lvl}
\end{figure}

We finish this section by expanding on the geometry of Cherenkov radiation and checking that the causality conditions are satisfied.
For the case where we have a single stationary point for a mode with group velocity $v_g$, we can find the critical frequency $\tilde{\omega}$ that solves (13). The phase speed at this frequency matches the charge velocity $v$ and results in the phase matching condition where path difference $CF=2BF\sin(\theta)$.
We can define the transit time $\tau$ of radiation reaching $(r,z)$ at time $t$ (as in Figure 6), where $\tilde{v}_r$ is the complex radial velocity or group velocity for wavenumber $k_{\perp}$:

\begin{align}
	\begin{split}
		\tilde{\tau}&=t-\dfrac{\sqrt{(z-v\tau)^2+r^2}}{\tilde{v}_g(\tilde{\omega})}	
	\end{split}.
\end{align}

The transit time $\tau$ connects the retarded time $t'=t_0+\tau$ and the charge position $vt'$ that is the source for the radiation at $(t,\mathbf{r})$. In the case of weak dissipation, the real group velocity will be a relevant physical quantity and the below causality condition must be satisfied:

\begin{align}
	t-\dfrac{z}{v}-r\sqrt{\dfrac{1}{v_g(\omega )^2}-\dfrac{1}{v^2}}\geq 0.
\end{align}

Defining the group cone (13) and the retarded current density ensures that this condition is automatically satisfied. To satisfy non-degeneracy, we require that the complex GVD not to be identically zero.

Care must be taken here, as the real group velocity defined by $v_g=\left(\Re\dfrac{dk}{d\omega}\right)^{-1}$ is not always a reliable measure of the signal velocity, especially in cases where we have significant dissipation and superluminal group velocities. For our purposes, we assume we're working with timescales where the group shape is still is coherent. In the next section, we'll consider the fields in the case of very small dissipation when we use the steepest descent method and the case of weak dissipation where we numerically solve for the fields.

\subsection{Cherenkov radiation in the EIT regime}
In general, the introduction of a nonzero loss term breaks the Cherenkov cone-like condition and dampens the field amplitude. For a large enough loss, the highly anisotropic character of the radiation and angular sensitivity to the charged particle's speed is weakened. 
Ideally we would want to get contributions close to the center of the transparency window, reduce the gaseous medium's particle density and introduce background refractive index enhancement to reduce $\Im{\chi}$. For the case of EIT with Rubidium D1 line spectra using fields with the linear polarization, we can select a transition between levels with magnetic number $m=0$ for EIT and introduce a static magnetic field to cause large enough energy level shifts for levels with the same nonzero $m$. For large enough detuning, these transitions will introduce a frequency-independent contribution to the susceptibility $\epsilon_{\infty}$.

We use the control scheme outlined in Sec. 3. The frequency range where Cherenkov radiation is transmitted is $[\omega_{31}+a,\omega_{31}+b]$, which contains all frequencies that contribute non-trivially to the radiation. We develop a scheme to evolve the fields corresponding to a certain mode $\omega_i$ after some time $\Delta t$. 

To define the time-evolving field, we first state the group velocity vector for Cherenkov radiation:

\begin{align}
	\mathbf{v}_{g, Ch}(\Delta\omega)=\dfrac{\beta^{-1}\hat{z}+n_{\perp}\hat{r}}{(1+\chi')+0.5(\omega_{31}+\Delta\omega)\dfrac{d\chi'}{d\Delta\omega}}
\end{align}

Suppose the wave propagates for small time $\Delta t$, such that the group profile remains coherent, from its initial point at $\{t,z,r_{\perp}\}$. Since the peak propagates with velocity $\mathbf{v}_{g, Ch}$, the new positions are given by $r'=r+(\mathbf{v}_{g, Ch}(\Delta\omega_i)\cdot\hat{r})\Delta t$, $z'=z+(\mathbf{v}_{g, Ch}(\Delta\omega_i)\cdot\hat{z})\Delta t$. We can express $\Delta t +z'-z=\alpha(vt-z)$ and $r'-r=\beta r$ and they satsify the cone condition if $\alpha=\beta$. This condition gives us the relation $\dfrac{v-\mathbf{v}_{g, Ch}(\Delta\omega_i)\cdot\hat{z}}{v\mathbf{v}_{g, Ch}(\Delta\omega_i)\cdot\hat{r}}=v_{g,r_{\perp}}^{-1}$, which can be obtained from (19). The cone condition, and the condition for non-degeneracy, is then automatically satisfied and we can define the time-evolving averaged Cherenkov Poynting vector, and the radiant power per unit length at a fixed radius in the axial direction $\zeta_z=2\pi r_{\perp}\expval{S}$, for modes $\tilde{\omega}_i\in \mathcal{S}_{t,z,r_{\perp}}$:

\begin{align}
	\begin{split}
		&\zeta_z(\mathbf{r}+\mathbf{v}_{g, Ch}(\Delta\tilde{\omega}_i)\Delta t,t+\Delta t)=\sum_{\tilde{\omega}_i\in \mathcal{S}_{t,z,r_{\perp}}}\pi(r+(\mathbf{v}_{g, Ch}(\Delta\omega_i)\cdot\hat{r})\Delta t)\\
		&\cross \mu^{-1}\Re{E_{r,\tilde{\omega}_i}(\mathbf{r},t)B^{*}_{\tilde{\omega}_i}(\mathbf{r},t)}e^{-2\Delta t\Im{\mathbf{k}\cdot\mathbf{v}_{g, Ch}(\Delta\tilde{\omega}_i)+\Delta\tilde{\omega}}}.
	\end{split}
\end{align}

The peak intensity of the radiation is confined to a thin cylinder about the z-axis because the dissipation and the spatial coherence depend only on the bandwidth and radial wavevector. For small $r$, the spatial phase $k_r'(\omega)r$ for different frequencies is small enough that the spectral components add coherently. The dissipation part $k_r''(\omega)r$'s shaping for the spectral profile is negligible for these distances. For large $r$, the phase differences are large enough that the phase is rapidly changing with frequency and the contributions to the field come from the critical points satisfying (13) or from nodes for the Gauss-Laguerre quadrature. The coherent addition of spectral components also depends on the retarded time through the relative spectral phase $\Delta\omega t'$, and hence on the radial components of group velocity and GVD. It follows that minimizing bandwidth and the range of $k_r$, for frequencies that contribute non-trivially to the radiant power, increases the coherence distance and time. The background susceptibility $\epsilon_{\infty}$ and the system+field parameters can be used to control the radiation spectrum and the radial refractive index through manipulation of the susceptibilities (4,5). EIT and the slow light condition allow for strong correlations between spectral components such that the fields due to multiple charges can stay in this coherence regime. With a significant enough accumulation of Cherenkov polaritons in the medium, we can expect a change in the optical response functions. In the next section, we consider the perturbative effects to the EIT state due to small fields.

\begin{figure}
	\hfill
	\subfigure[$\Re n_r$]{\includegraphics[width=5cm]{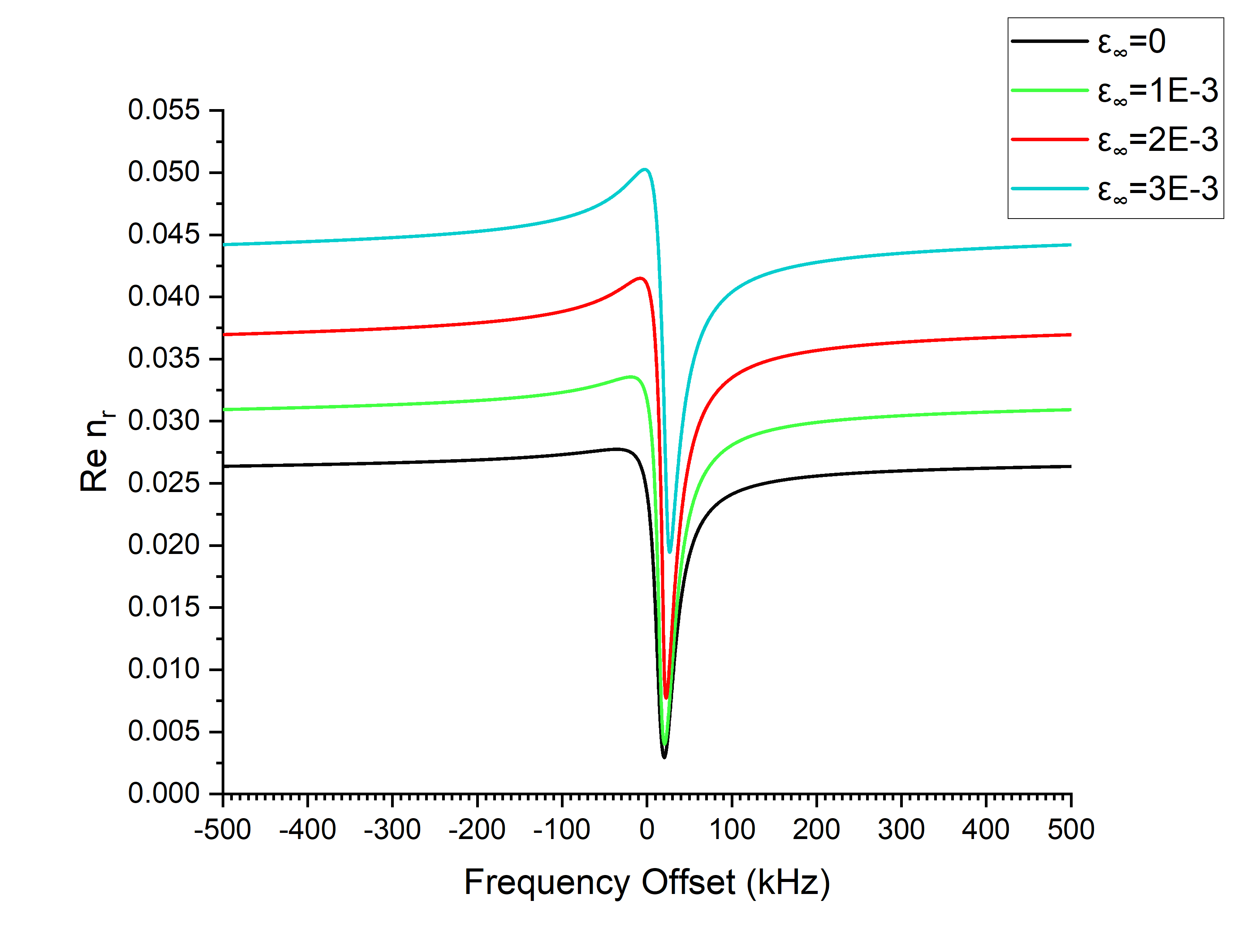}}
	\hfill
	\subfigure[$\Im n_r$]{\includegraphics[width=5cm]{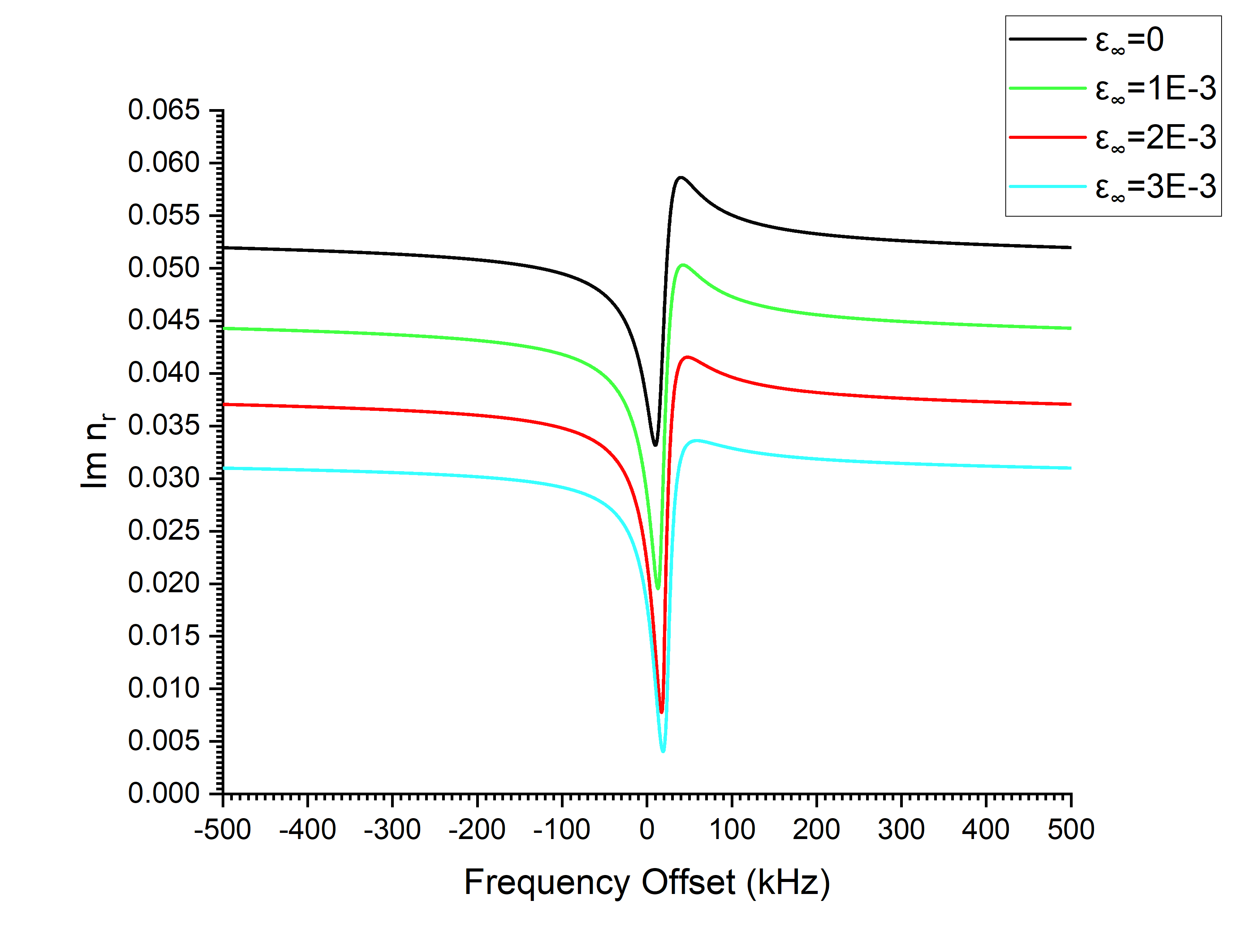}}
	\hfill
	\caption{Plots showing the radial refractive index $n_r$ for various values of $\epsilon_{\infty}$. The decrease of $\Im n_r$ for large $\epsilon_{\infty}$ proportionally decreases the dissipation and is accompanied by increased and sharper dispersion. }
	\label{fig:5lvl}
\end{figure}

\begin{figure}
	\hfill
	\subfigure[$\Omega_{2}=1/35 \gamma_{31}$]{\includegraphics[width=5cm]{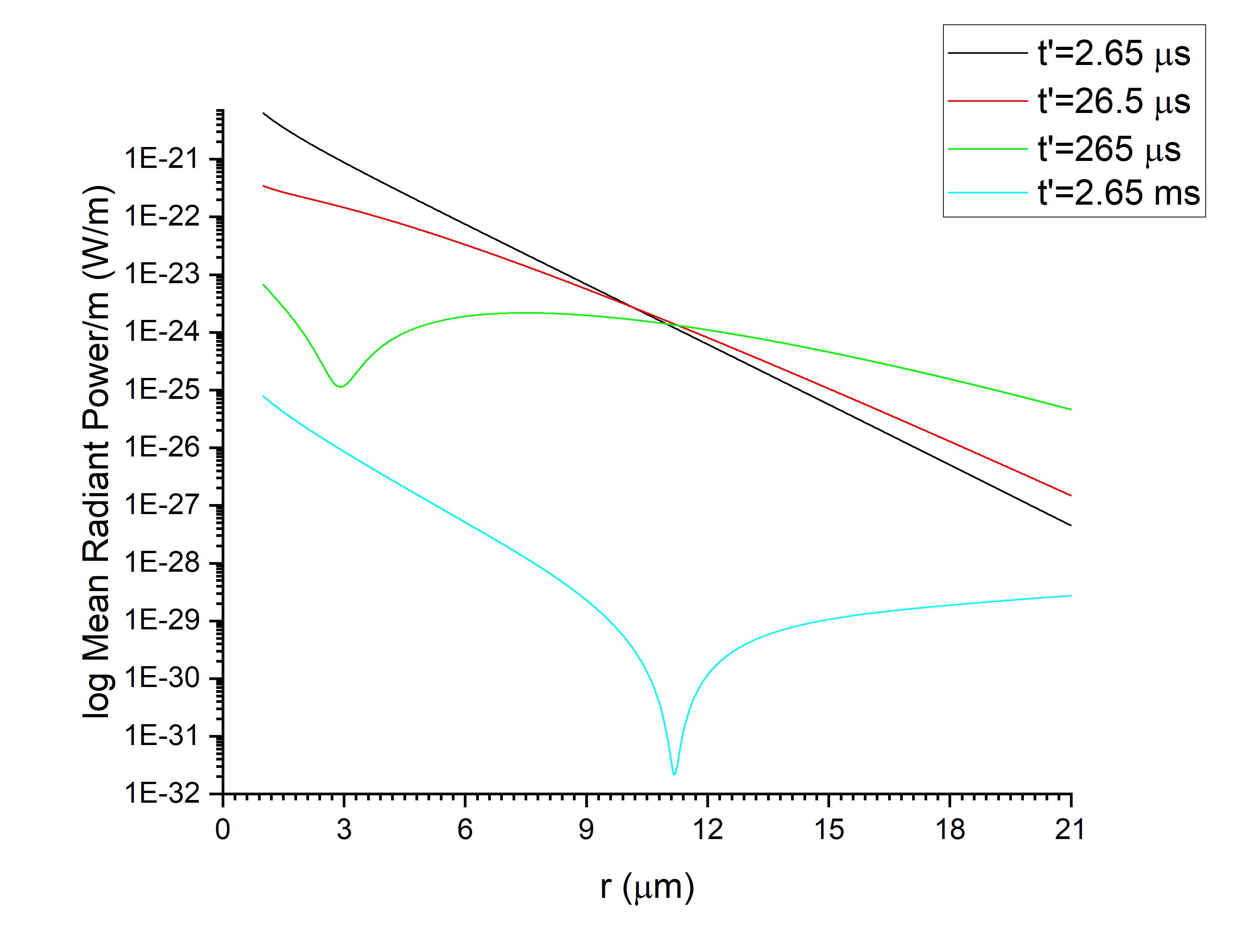}}
	\hfill
	\subfigure[$\Omega_{2}=2/35 \gamma_{31}$]{\includegraphics[width=5cm]{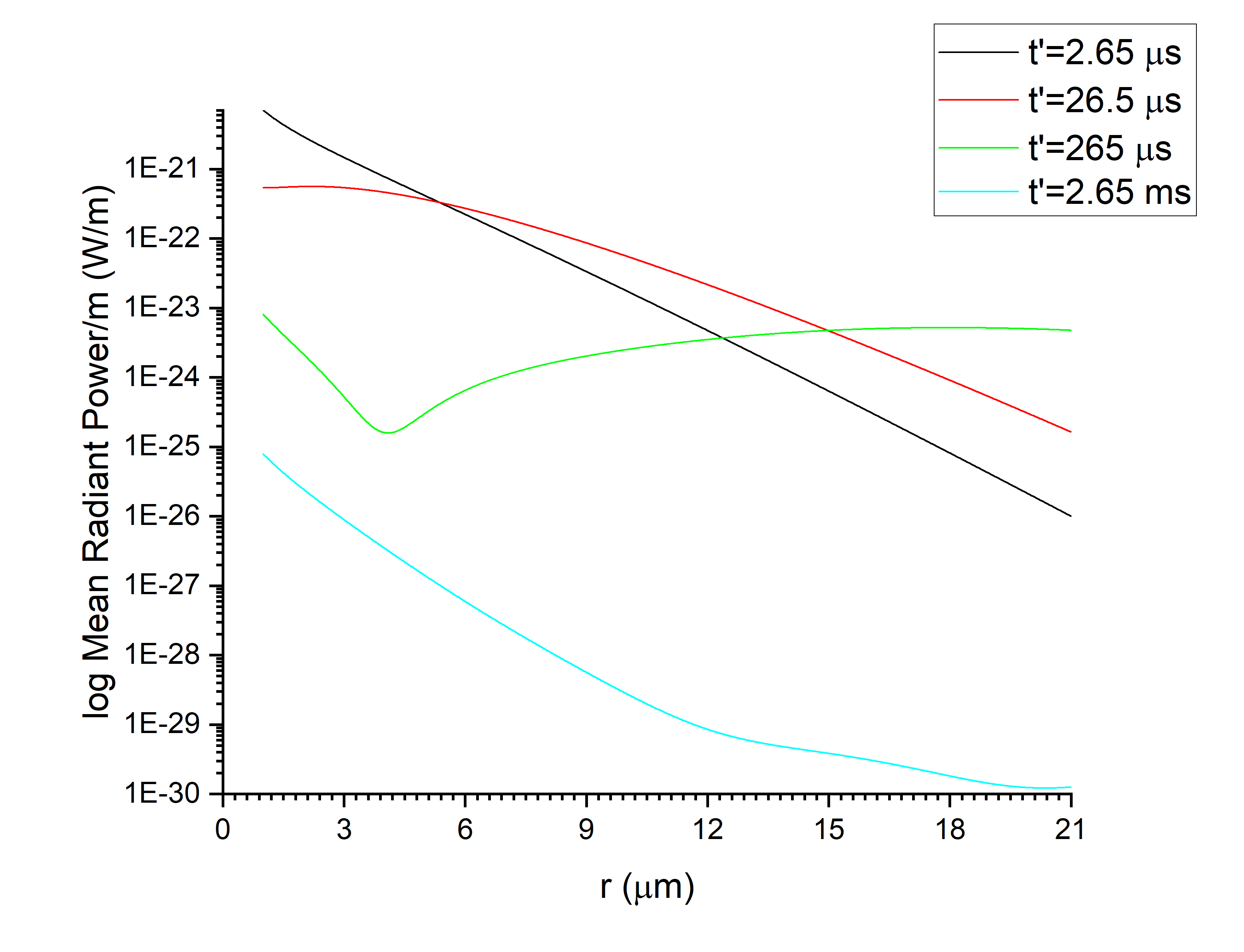}}
	\hfill
	\caption{Plots for $\zeta_z(\mathbf{r},t)$ with varying $r$ for different times. }
	\label{fig:5lvl}
\end{figure}
\newpage

\subsection{Perturbation to steady states: Developing a control method to detect high energy particles}
We can consider a couple of approaches in detecting the presence of Cherenkov radiation. Using a sensor is a viable approach but we can also use the methods of atomic spectroscopy and interferometry by tuning the optical properties of the medium and its sensitivity to small perturbations.
\newline

Consider a setup of atoms in an EIT steady state exposed to Cherenkov radiation described by the Lindbladian $\mathcal{L_\mathcal{J}}$. We assume that $\mathcal{L_\mathcal{J}}$ has a unique steady state, with projection operator on the steady state subspace given by $\mathcal{P}_{\infty}=\sum_{\lambda_k}\ket{\psi_{\lambda_k}}\bra{\psi_{\lambda_k}}$, and introduce the perturbation operator $g(t)\mathcal{O}$ (corresponding to $-i\comm{V}{\cdot}$ in the field interaction picture) to describe the Cherenkov radiation-atom interaction. The time evolution from an initial state is given by $\mathcal{T}e^{\int_{-\infty}^{t}d\tau(\mathcal{L_\mathcal{J}}+g(t)\mathcal{O})}\ket{\rho_s}$. We assume that the perturbation due to Cherenkov radiation is switched on at $t=0$, $g(t)$ is slowly time-varying, and use the Kubo formula to solve for the first-order perturbation \cite{lindblad}: 

\begin{align}
\mathcal{T}_t^{(1)}\ket{\rho_s}=\int_{-\infty}^{t}d\tau g(\tau)
e^{(t-\tau)\mathcal{L_\mathcal{J}}}\mathcal{O}\ket{\rho_s}
\end{align}

In the case where $g(t)=G\Theta(t)$, $\mathcal{T}_t^{(1)}\ket{\rho_s}=G\mathcal{L_\mathcal{J}}^{-1}\mathcal{O}\ket{\rho_s}$. The above equation assumes we start from the steady state for a particular choice of parameters. We then apply the perturbation to the system and calculate the first order perturbation to the steady state using the Kubo formula. The first order perturbation on a steady state describes a leakage from the asymptotic state subspace \cite{lindblad}. For the choice of parameters $\Omega_1=0.1 \text{ MHz}$, $\Omega_2=1.0 \text{ MHz}$, $\Delta_{1}=\Delta_{2}=0$, the Lindbladian is diagonalizable with 1 physical state and 8 unphysical traceless states with negative real part of eigenvalues $\lambda_k$, $\left( \dfrac{\Omega_{1}^2+\Omega_{2}^2}{\gamma_{31}}\leq|\lambda_k|\leq\gamma_{31}\right)$. This dissipative energy gap, $\Delta_{dg}$, is very similar to the curve of $\dfrac{\Omega_{1}^2+\Omega_{2}^2}{\gamma_{31}}$. Increasing the dissipation gap to large values $\Delta_{dg}$ will decrease the leakage from the asymptotic subspace. We observe that small fields lead to almost no change in the coherence for our chosen probe field as contributions due to dephasing negates any population of the other states. For large Rabi frequencies, we observe that the perturbation magnitude decreases, as the field strength increases $\Delta_{dg}$ and $\Gamma'$. However, as Figure 9 shows, this effect is somewhat offset by choosing large $\abs{\Delta_{2}}$. The perturbative effect seems to be strongest between the two sub-bands (where the perturbation is minimized) and to the right of the second sub-band. There are trade-offs in choosing smaller Rabi frequencies in having to deal with larger absorptivity, while larger Rabi frequencies will broaden the optical response lineshapes. 

Figure 9 show regions in parameter spaces in which we can get significant change to $\rho_{31}$, whilst retaining choice for the real and imaginary parts, for small perturbations. In determining the full perturbation to the steady state, we can configure the initial system state of the detection system towards a desired change in the optical response functions $n',n''$. For example, we could be in the region with maximized GVD and the perturbed state will modify the chirp spectrum of the polarized field of the Cherenkov radiation or another probe field and cause a delay in comparison to the pulses propagating in the unperturbed system. Small perturbations can also be very useful if the peak frequencies of the Cherenkov field or of another probe pulse are near the anomalous dispersion region, such that the perturbative response will lead to phenomena such as superluminal group velocities. Alternatively we could measure a different propagation angle for Cherenkov radiation due to change in $\eta$. This method allow us to observe the change in group profile/chirp spectrum of the Cherenkov radiation or a sequence of probe pulses. 

\begin{figure}
	\hfill
	\subfigure[$\Re\mathcal{T}_t^{(1)}\rho_{13,s}$]{\includegraphics[width=5cm]{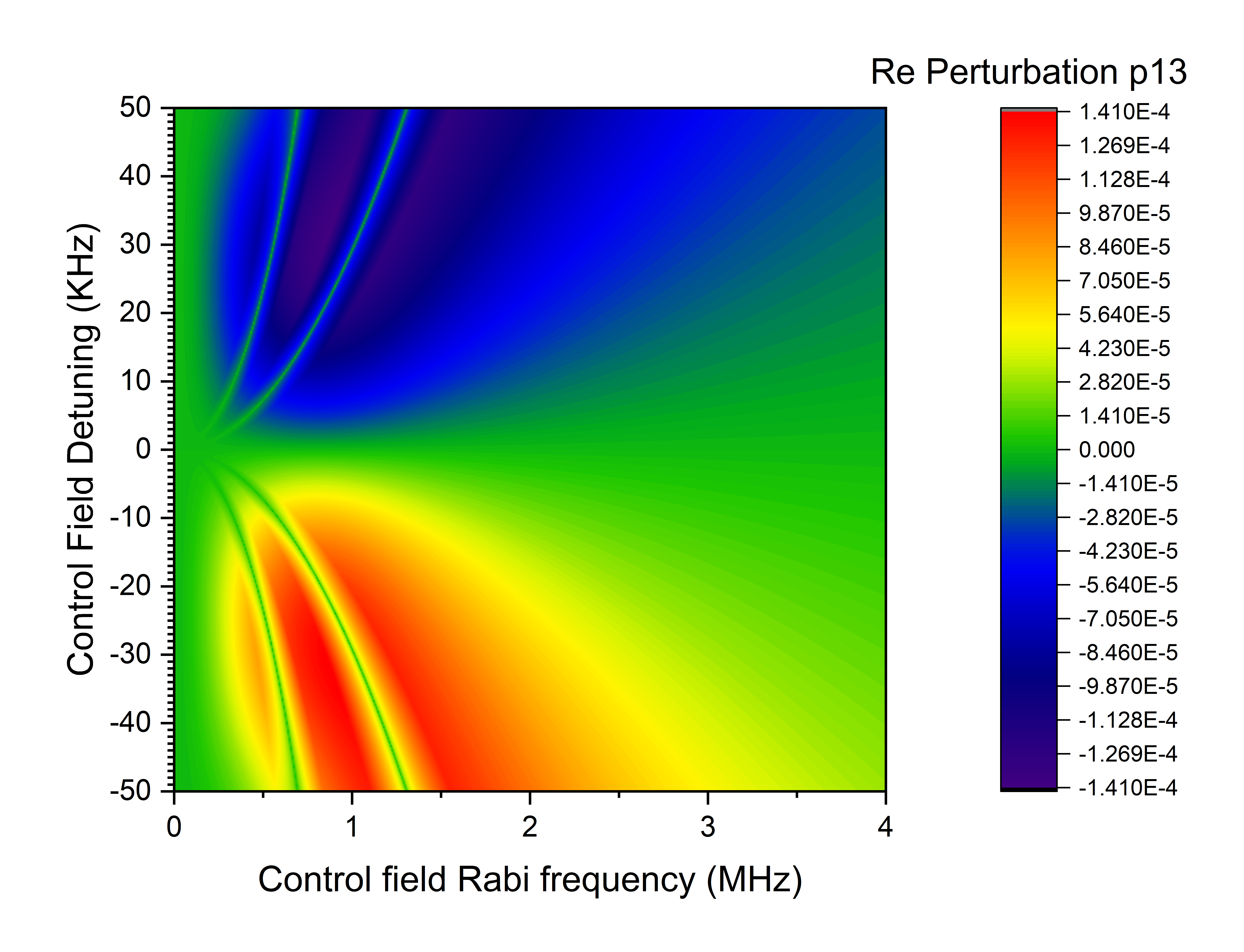}}
	\hfill
	\subfigure[$\Im\mathcal{T}_t^{(1)}\rho_{13,s}$]{\includegraphics[width=5cm]{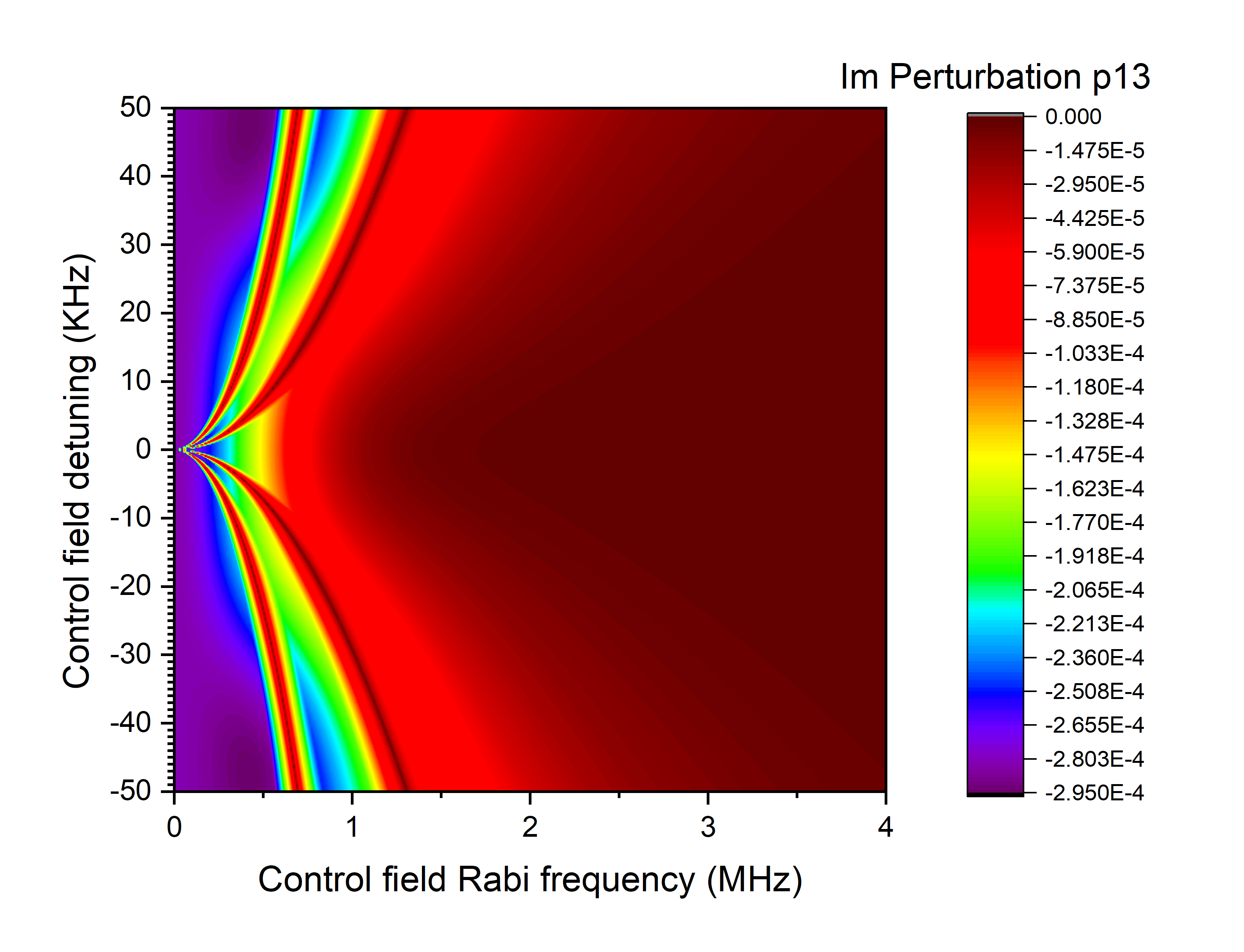}}
	\hfill
	\caption{Plots showing the real and imaginary parts of the perturbation $\mathcal{T}_t^{(1)}\rho_{13,s}$, due to perturbation $G\mathcal{O}$ after 1 $\mu$s (color axis). $\Omega_1=0.1\Omega_2$, $\gamma_{12}=4.3E-5$ , $\Delta_2=-5.7E-4$, $\Omega_{pert}=2.7E-4$.}
	\label{pert2}
\end{figure}
\newpage

\subsubsection{Control method summary}
We explicitly state our control protocol for detection of high energy particles. Our detection scheme is developed for two different systems of interest with different objectives for control. The first is for the accumulation of Cherenkov polaritons, it prioritises high transmission, nonlinear modes and slow group velocity to increase the yield of radiation. The second system is tailored towards the detection of the Cherenkov radiation, involving choosing atomic and field parameters such that the optical response functions in the medium will develop a desired change in properties such as the chirp spectra of a probe pulse post-medium. 
We start from the same base conditions as those listed in Sec 3.1.
\newline

For the first system, we note that the properties of the Cherenkov radiation are determined by the parameters $\omega_{31}$, $\mathbf{r_{\perp}}$, $\sigma$, $\eta$, $\phi$, $c_0$ and the bandwidth $[a,b]$ (from Sec 4.2). We set the cutoff speed to be equal to $\beta$, choose a Cherenkov parameter set to give us a desired form for the radiation, choose the system+field parameters to shape the radial wavenumber function $k_r(\omega)$ and determine the frequency interval $[\omega_{L1}+a,\omega_{L1}+b]$ (with $a,b$ being the solutions to $1-\beta^{-2}+\Re{\chi}=0$ in the window centered at the peak frequency). The bandwidth for the coherent group mode is limited by the EIT linewidth $\Gamma'$ and adjusting the control Rabi frequency $\Omega_2$ is one way to broaden the linewidth, decrease dissipation and increase coherence times. The axial and radial group velocities are determined using (22) and (6) 
which can then be used to find any critical points satisfying (13) and then get the radiant power density (23). We can then optimise the power density for a train of high energy particles associated with some distribution of arrival times.

For the second system, we optimise the detection process by choosing the set of system+field parameters such that the initial state is perturbed by the Cherenkov field to a desired state, which has different optical responses. We determine the regions in the plot of $\mathcal{T}^{(1)}(\rho_s)$ in the system+field parameter space which correspond to perturbed states we target. For example, we can choose to have the initial state to cause high GVD for a probe pulse and thus a significant enough perturbation that we get a measurable chirp and pulse delay for propagation in the perturbed medium.
\newline

This concludes our treatment for detecting high energy particles using EIT in atomic media. In the final section, we review our results and consider their extension to more complex systems.
\newpage

\section{Summary and vision for nanophotonics}
We derived the optical response functions for the three-level $\Lambda$ system driven to a EIT state in an open weakly coupled system using the Lindblad formalism to consider the effects of dephasing and decoherence after determining the field interaction Hamiltonian. We used the Bloch vector formalism to transition from the microscopic to the macroscopic picture where we discussed the optical response functions (dispersion and absorptivity), and showed how an EIT state with a choice of parameters would lead to a narrow transparency window with strong non-linear dispersion effects that created phenomena such as slow light and anomalous dispersion.

This led to our main focus of this chapter which is on developing quantum control schemes to use EIT's optical response and sensitivity as a spectroscopic tool to detect phenomena such as the passage of high-energy particles. Our model was developed in the regime of weak fields with ultranarrow Fano resonances that capitalize on the properties of slow light and dissipation-free transmission. The group velocity dispersion (GVD) and features of the group velocity (such as in the presence of anomalous dispersion) (Sec 3) were considered in this regard.  

We discussed an application of our approach to the detection of high-energy charged particles through the accumulative stimulation of Cherenkov radiation, by proxy of the radiation behaving as slow moving Cherenkov polaritons in a medium undergoing EIT (Sec 4), and derived a weak-field approximation analytical model that allows us to calculate the group profile and its time-dependence. We then considered how small perturbations would affect the steady state configuration determining the leakage from the steady-state subspace, the first order perturbation and the dissipation gap that is a measure of the minimum decay rate of contributions from states outside the steady state subspace. In tandem with the known optical response functions, we showed that we can realize a control scheme that optimises the sensitivity of the optical response functions to the perturbation.

We note that whilst we spend most of our attention on the atomic and field properties, the high energy particles can also be controlled in a similar manner to the atomic transitions. Work in the field of PINEM (Photon induced near-field electron microscopy) \cite{PINEM} shows that high energy charged particle wavefunctions can be modulated over nanostructures with fringing evanescent fields and enhance probing of a train of randomly arriving high energy particles , introduce correlations between the particles, and increase the perturbation to a bound electron state \cite{Freel}. We can control the Rabi phase for a random arrival process and increase coherent population build up in comparison to a train of uncorrelated wavefunctions. Further approaches include reverse engineering the high energy particle wavefunction from the photon spectral density function as well as through optically probing the atomic coherences \cite{Cohlight}.

Whilst this chapter focused on a particular atomic system, the results presented here would have similar analogues in more complex systems, such as photonic crystals and nanostructures, providing a strong relevance to quantum information theory and optical storage. There is vast literature in research of EIT in nanostructures \cite{Polariton}. Coupling of the EM field to material modes in the system leads to formations of quasiparticles similar to dark state atomic EIT polaritons. Examples already researched include surface plasmon polaritons, in the case of introducing a substrate below the dielectric, or quasiparticles that form in bulk such as exciton-polaritons.

Surface plasmon polaritons (SPPs) show powerful applications in the production of tunable power-enhanced Cherenkov radiation. In \cite{GrapheneCH}, a dielectric substrate with a buffer layer deposited with multiple layers of graphene was used to enhance the production of Cherenkov radiation due to an electron beam by coupling the radiation to the SPP modes, resulted in increasing the intensity of the radiation by more than 2 orders of magnitude. The properties of the buffer, substrate and SPP modes can be used to control the intensity increase and the radiation frequency. Of particular interest is the analogous phenomenon of plasmon-induced transparency (PIT) realized in metamaterials consisting of two plasmonic resonators coupled to a waveguide \cite{EITPM}. One scheme utilizes coupling between a dark and bright resonator to achieve a transparency window and optical response functions similar to that of EIT in atomic media. Graphene metamaterials have seen wide applications in this field due to its low-loss, shiftable Fermi energies and electronic properties.
A number of different geometrical approaches have theoretically showed PIT \cite{EITtp, EITtp2} in the THz regime and this phenomena's applicability to quantum control.  Due to the geometry of the resonators, spacings in the metamaterial, and properties of the graphene layer give high controllability for the electromagnetic response. The features of enhanced slow light and trapping as well as the amplification of signals and the presence of PIT suggests a possible extension for this chapter's results.  

Exciton interactions are another area where this research can be extended. The strong interaction of excitons with matter and light, the capability to induce EIT, and the formation of strongly non-linear exciton-polariton states \cite{GiantRyd} suggest a future direction for sensing high energy particles with EIT in more complex media. One particular application is for systems with Rydberg excitonic states, which yield significantly higher nonlinearities in the optical response and few photon strongly correlated states that result from the Rydberg blockade effect \cite{GiantRyd}. With the introduction of the blockade effect, EIT is disrupted and the nonlinearities are transferred to the few photon states. Benefits of systems with Rydberg excitonic states include greatly increased interaction strengths and applications of ultralow field intensities causing reduced decoherence and line broadening effects. With the traditional features of EIT including the transparency window and slow light along with the strongly correlated few photon states and quantum control aspect, there is good scope in detecting high energy particles through the production and control of slow moving Cherenkov polariton-excitons.

\clearpage
\bibliography{bib2.bib}
\bibliographystyle{ieeetr}

\end{document}